\documentclass[11pt,a4paper]{article}
\pdfoutput=1
\usepackage[utf8]{inputenc}
\usepackage{a4wide}
\usepackage{amsmath}
\usepackage{amssymb}
\usepackage{amsfonts}
\usepackage{cite}
\usepackage{color}
\usepackage{adjustbox}
\usepackage{multirow}
\usepackage{pdflscape}
\usepackage{textcomp}
\usepackage{gensymb}
\usepackage{graphicx}
\graphicspath{{figures/}}
\usepackage{diagbox}
\usepackage{pifont}
\usepackage{bigstrut}
\usepackage[small,bf]{caption}
\usepackage{tabulary}
\usepackage{booktabs}
\usepackage{stackengine}
\usepackage{mathtools}
\usepackage{enumerate}
\usepackage[unicode]{hyperref}

\newcommand{\be}{\begin{equation}}
\newcommand{\ee}{\end{equation}}

\def\1{\mathbf{1}}
\def\op{\mathbf{1^\prime}}
\def\2{\mathbf{2}}
\def\3{\mathbf{3}}
\def\tp{\mathbf{3^\prime}}
\def\a{\alpha}
\def\b{\beta}
\def\g{\gamma}
\def\G{\Gamma}

\def\D{\Delta}

\def\L{\Lambda}
\def\l{\lambda}

\def\s{\sigma}
\def\t{\tau}
\def\th{\theta}

\def\o{\omega}

\def\ot{\otimes}

\def\di{{\rm d}}
\def\Ec{E^c}
\def\Nc{N^c}
\def\Hd{H_d}
\def\Hu{H_u}
\def\vd{v_d}
\def\vu{v_u}
\def\y{\mathcal{Y}}
\def\Lp{\Lambda^\prime}
\def\Lpp{\Lambda^{\prime\prime}}
\def\Lppp{\Lambda^{\prime\prime\prime}}
\def\gp{g^\prime}
\def\gpp{g^{\prime\prime}}
\def\gppp{g^{\prime\prime\prime}}
\DeclareMathOperator{\tr}{tr}
\DeclareMathOperator{\rank}{rank}
\DeclareMathOperator{\diag}{diag}

\DeclareMathOperator{\im}{Im}
\DeclareMathOperator{\re}{Re}

\DeclarePairedDelimiter{\vevdel}{\langle}{\rangle}
\newcommand{\vev}{\vevdel}
\allowdisplaybreaks
\numberwithin{equation}{section}
\makeatletter
\g@addto@macro\bfseries{\boldmath}
\makeatother
\renewcommand{\star}{*}

\begin{document}

\begin{titlepage}

\vspace*{-15mm}
\begin{flushright}
SISSA 47/2018/FISI \\ 
IPMU18-0187 \\
IPPP/18/98
\end{flushright}
\vspace*{8mm}

\begin{center}
{\bf\LARGE Modular $S_4$ Models of Lepton Masses and Mixing}\\[8mm]
P.~P.~Novichkov$^{\,a,}$\footnote{E-mail: \texttt{pavel.novichkov@sissa.it}},
J.~T.~Penedo$^{\,a,}$\footnote{E-mail: \texttt{joao.penedo@sissa.it}},
S.~T.~Petcov$^{\,a,b,}$\footnote{Also at Institute of Nuclear Research and Nuclear Energy, Bulgarian Academy of Sciences, 1784 Sofia, Bulgaria.}, 
A.~V.~Titov$^{\,c,}$\footnote{E-mail: \texttt{arsenii.titov@durham.ac.uk}} \\
\vspace{8mm}
$^{a}$\,{\it SISSA/INFN, Via Bonomea 265, 34136 Trieste, Italy} \\
\vspace{2mm}
$^{b}$\,{\it Kavli IPMU (WPI), University of Tokyo, 
5-1-5 Kashiwanoha, 277-8583 Kashiwa, Japan} \\
\vspace{2mm}
$^{c}$\,{\it Institute for Particle Physics Phenomenology, 
Department of Physics, Durham University,\\ 
South Road, Durham DH1 3LE, United Kingdom}
\end{center}
\vspace{8mm}

\begin{abstract}
\noindent 
We investigate models of charged lepton and neutrino 
masses and lepton mixing 
based on broken modular symmetry.
The matter fields in these models 
 are assumed to transform in irreducible representations of
the finite modular group $\Gamma_4 \simeq S_4$.
We analyse the minimal scenario 
in which the only source of symmetry breaking
is the vacuum expectation value of the modulus 
field. In this scenario there is no need to 
introduce flavon fields. Using the basis for 
the lowest weight modular forms found earlier,
we build minimal phenomenologically viable models
in which the neutrino masses are generated via 
the  type I seesaw mechanism. While 
successfully accommodating charged lepton masses,
 neutrino mixing angles and mass-squared differences,
these models predict the values of the lightest neutrino mass 
(i.e., the absolute neutrino mass scale), of the Dirac and 
Majorana CP violation (CPV) phases, as well as  
specific correlations between
the values of the atmospheric neutrino mixing parameter 
$\sin^2\theta_{23}$ and i) the Dirac CPV phase $\delta$,
ii)~the sum of the neutrino masses, and 
iii) the effective Majorana mass in neutrinoless double beta decay.
We consider also the case of residual symmetries
$\mathbb{Z}^{ST}_3$ and $\mathbb{Z}^S_2$ respectively 
in the charged lepton and neutrino sectors, 
corresponding to specific vacuum expectation values of the modulus.
\end{abstract}
\end{titlepage}
\setcounter{footnote}{0}

\section{Introduction}
\label{sec:intro}
Understanding the origin of the flavour 
structure of quarks and leptons 
continues to be a highly challenging problem. 
Adding to this problem is the pattern of two large and one small 
mixing angles in the lepton sector, revealed by the data 
obtained in neutrino oscillation experiments 
(see, e.g.,~\cite{PDG2018}).
The results of the recent global analyses 
of these data show also that a neutrino mass spectrum with 
normal ordering (NO) is favoured over the spectrum with inverted 
ordering (IO), as well as a preference for a value of the 
Dirac CP violation (CPV) phase $\delta$
close to $3\pi/2$ (see, e.g.,~\cite{Capozzi:2018ubv}).

 The observed 3-neutrino mixing pattern 
can naturally be explained by extending 
the Standard Model (SM) with a flavour symmetry 
corresponding to a non-Abelian discrete (finite) group $G_f$ 
(see, e.g.,~%
\cite{Altarelli:2010gt,Tanimoto:2015nfa,King:2013eh,Petcov:2017ggy}).
This symmetry is supposed to exist at some high-energy scale 
and to be broken at lower energies to residual symmetries of   
the charged lepton and neutrino sectors.
Extensive studies  of the non-Abelian discrete flavour 
symmetry approach to the (lepton) flavour problem 
have revealed that, typically, the breaking of 
the flavour symmetry requires the introduction of a 
large number of scalar fields (flavons). 
These fields have to develop a set of particularly 
aligned vacuum expectation values (VEVs). Arranging 
for such an alignment requires in turn 
the construction of rather elaborate scalar potentials.

  A new and very interesting approach to the lepton flavour
problem, based on invariance under 
the modular group, has been proposed in Ref.~\cite{Feruglio:2017spp}  
where also models based on the 
finite modular group $\G_3 \simeq A_4$
have been constructed. Although the models found in
Ref.~\cite{Feruglio:2017spp} were not realistic and made use 
of a minimal set of flavon fields, this work 
inspired further studies of the modular invariance 
approach to the lepton flavour problem.
In Ref.~\cite{Kobayashi:2018vbk} a realistic model
with modular $\G_2 \simeq S_3$ symmetry 
was built with the help of a minimal set 
of flavon fields. In the most economical versions 
of the models with modular symmetry, 
the VEV of the modulus $\tau$ can be, in principle, 
the only source of symmetry breaking 
without the need of flavon fields.
A realistic model of the charged lepton 
and neutrino masses and of neutrino mixing without 
flavons, in which the modular $\G_4 \simeq S_4$ 
symmetry was used, was constructed in~\cite{Penedo:2018nmg}.
Subsequently, lepton
flavour models with and without flavons based on 
the modular symmetry $\G_3 \simeq A_4$ 
have been proposed in Refs.~%
\cite{Criado:2018thu,Kobayashi:2018scp}. 
 
 In the present article, building on the 
results obtained in Ref.~\cite{Penedo:2018nmg},
we construct in a systematic way flavour models 
based on the finite modular group $\G_4\simeq S_4$ and study 
in detail their phenomenology. We focus on the case when the 
light neutrino masses are generated via the 
type I seesaw mechanism and where no flavons are introduced.

 The article is organised as follows.
In Section~\ref{sec:framework}, we briefly describe the 
modular symmetry approach to lepton masses and mixing 
proposed in Ref.~\cite{Feruglio:2017spp}.
In Section~\ref{sec:seesaw}, we construct minimal modular-invariant
seesaw models.
In Section~\ref{sec:numerical}, we perform a thorough numerical analysis, identify viable models and study their phenomenology.
In Section~\ref{sec:ressymm}, we discuss the implications of preserving residual symmetries of the modular group, 
while in Section~\ref{sec:corrections} we discuss potential sources of corrections. 
Finally, in Section~\ref{sec:summary} we summarise our conclusions.

\section{The Framework}
\label{sec:framework}

\subsection{Modular group and modular forms}
\label{subsec:modinv}
 The modular group $\overline{\G}$  
is the group of linear fractional transformations $\g$ 
acting on the complex variable $\t$ belonging to 
the upper-half complex plane as follows:
\be
\g\t = \frac{a\t + b}{c\t + d}\,,
\quad
\text{where} 
\quad
a,b,c,d \in \mathbb{Z}
\quad
\text{and}
\quad
ad - bc = 1\,,~~{\rm Im}\tau > 0\,.
\label{eq:linfractransform}
\ee
%
Since changing the sign of $a,b,c,d$ simultaneously does not change eq.~\eqref{eq:linfractransform}, 
the group $\overline{\G}$ is isomorphic to the projective special linear group 
$PSL(2,\mathbb{Z}) = SL(2,\mathbb{Z})/\mathbb{Z}_2$, 
where $SL(2,\mathbb{Z})$ is the group
of $2\times2$ matrices with integer elements and unit determinant, 
and $\mathbb{Z}_2 = \{I, -I\}$ is its centre ($I$ being the identity element).
The modular group is generated by two transformations $S$ and $T$ satisfying
\be
S^2 = \left(ST\right)^3 = I\,.
\ee
Representing these transformations as
\be
S = \begin{pmatrix}
0 & 1 \\
-1 & 0
\end{pmatrix}\,,
\qquad
T = \begin{pmatrix} 
1 & 1 \\
0 & 1
\end{pmatrix}\,,
\ee
%
we obtain
\be
\t \xrightarrow{S} -\frac{1}{\t}\,, 
\qquad
\t \xrightarrow{T} \t + 1\,.
\ee
%

 Consider now the series of infinite normal subgroups 
$\G(N)$, $N = 2,3,4,\dots$, 
of $SL(2,\mathbb{Z})$ given by
\be
\G(N) = \left\{
\begin{pmatrix}
a & b \\
c & d
\end{pmatrix} 
\in SL(2,\mathbb{Z})\,, 
\quad
\begin{pmatrix}
a & b \\
c & d
\end{pmatrix} =
\begin{pmatrix}
1 & 0 \\
0 & 1
\end{pmatrix} 
~~(\text{mod } N)
\right\}\,.
\ee
%
For $N=2$ we define $\overline{\G}(2) \equiv \G(2)/\{I,-I\}$, 
while for $N > 2$, since the element $-I$ does not belong to $\G(N)$, we have 
$\overline{\G}(N) \equiv \G(N)$.
The elements of $\overline{\G}(N)$ are in a one-to-one correspondence with 
the associated linear fractional transformations. 
The groups $\overline{\G}(N)$ are referred to as 
principal congruence subgroups of the modular group.
Taking the quotient $\G_N \equiv \overline{\G}/\overline{\G}(N)$, one obtains a finite modular group. Remarkably, for $N \leq 5$ the finite modular groups are isomorphic to 
permutation groups widely used in lepton flavour model building 
(see, e.g., \cite{deAdelhartToorop:2011re}). 
Namely, $\G_2 \simeq S_3$, $\G_3 \simeq A_4$, $\G_4 \simeq S_4$ and $\G_5 \simeq A_5$.

 Modular forms of weight $k$ and level $N$ are holomorphic functions $f(\t)$ 
transforming under the action of $\overline{\G}(N)$ in the following way:
\be
f\left(\g\t\right) = \left(c\t + d\right)^k f(\t)\,, 
\quad 
\g \in \overline{\G}(N)\,.
\ee
%
Here $k$ is even and non-negative, and $N$ is natural
  (note that $\G(1) \simeq SL(2,\mathbb{Z})$ and
  $\overline{\G}(1) \equiv \overline{\G}$).
Modular forms of weight $k$ and level $N$ form a linear space
of finite dimension.
It is possible to choose a basis in this space such that 
a transformation of a set of modular forms $f_i(\t)$ is described 
by a unitary representation $\rho$ of the finite modular group $\G_N$:
\be
f_i\left(\g\t\right) = \left(c\t + d\right)^k \rho\left(\g\right)_{ij} f_j(\t)\,, 
\quad 
\g \in \overline{\G}\,.
\label{eq:vvmodforms}
\ee
%
This result is the foundation stone of the approach to lepton masses and mixing
proposed in Ref.~\cite{Feruglio:2017spp}. 

 In the case of $N=2$, the modular forms of lowest 
non-trivial weight 2 form a two-dimensional linear space. 
One can find a basis in which the two generating modular forms 
are transformed according to the 2-dimensional irreducible representation (irrep) 
of $S_3$ \cite{Kobayashi:2018vbk}. In the case of $N=3$, 
the corresponding space has dimension 3, and the generating modular forms 
have been shown to form the triplet of $A_4$ \cite{Feruglio:2017spp}. 
For $N=4$, there are 5 linearly independent modular forms of weight 2. 
They are organised in a doublet and a triplet ($\tp$) of $S_4$ \cite{Penedo:2018nmg}.
Modular forms of higher weights $(k>2)$ can be constructed from 
homogeneous polynomials in the generating modular forms of weight 2.

\subsection{Modular-invariant supersymmetric action}
\label{subsec:modinvSUSYaction}
 In the case of $\mathcal{N} = 1$ rigid supersymmetry (SUSY), 
the matter action $\mathcal{S}$ reads
\be
\mathcal{S} = \int \di^4x\, \di^2\th\, \di^2\overline{\th}~ 
K(\t, \overline{\t}, \chi, \overline{\chi}) + 
\int \di^4x\, \di^2\th~W(\t, \chi) + 
\int \di^4x\, \di^2\overline{\th}~\overline{W}(\overline{\t}, \overline{\chi})\,,
\ee
%
where $K$ is the Kähler potential, $W$ is the superpotential 
and $\chi$ denotes a set of chiral supermultiplets contained in the theory 
in addition to the modulus $\t$. 
The integration goes over both space-time coordinates $x$ 
and Graßmann variables $\th$ and $\overline{\th}$.
The supermultiplets $\chi$ are divided into 
several sectors $\chi_I$. Each sector in general contains several supermultiplets.

 The modular group $\overline{\G}$ acts on 
$\t$ and $\chi_I$ in a specific way \cite{Ferrara:1989bc,Ferrara:1989qb}.
Assuming that the supermultiplets $\chi_I$ transform also in a certain 
representation $\rho_I$ of a finite modular group $\G_N$, 
we have
\be
\begin{cases}
\t \rightarrow \dfrac{a\t + b}{c\t + d}\,, \\[4mm]
\chi_I \rightarrow \left(c\t + d\right)^{-k_I} \rho_I(\g)\, \chi_I\,.
\end{cases}
\label{eq:modtransforms}
\ee
%
The transformation law for the supermultiplets $\chi_I$ is similar to that 
in eq.~\eqref{eq:vvmodforms}. However, $\chi_I$ are not modular forms, 
and thus, the weight $(- k_I)$ is not restricted to be an even non-negative number.
The invariance of $\mathcal{S}$ under the transformations 
given in eq.~\eqref{eq:modtransforms} requires the invariance of the superpotential $W$, 
while the Kähler potential $K$ is allowed to change by a Kähler transformation, i.e.,
\be
\begin{cases}
W(\t, \chi) \rightarrow W(\t,\chi)\,, \\[4mm]
K(\t, \overline{\t}, \chi, \overline{\chi}) \rightarrow K(\t, \overline{\t}, \chi, \overline{\chi}) 
+ f_K(\t,\chi) + \overline{f_K}(\overline{\t},\overline{\chi})\,.
\end{cases}
\ee
%
An example of Kähler potential which satisfies this requirement is given by 
\be
K(\t, \overline{\t}, \chi, \overline{\chi}) = - \L_0^2\,\log(-i\t + i\overline{\t}) 
+ \sum_{I} \frac{|\chi_I|^2}{(-i\t + i\overline{\t})^{k_I}}\,,
\label{eq:Kahler}
\ee
%
where $\L_0$ is a parameter with mass dimension one.%
\footnote{Note that we consider $\t$ to be a dimensionless chiral supermultiplet, 
as it is done in Ref.~\cite{Feruglio:2017spp}.}
Expanding the superpotential in powers of $\chi_I$, we have
\be
W(\t, \chi) = \sum_{n} \sum_{\{I_1,\dots,I_n\}}
\left(Y_{I_1\,\dots\,I_n}(\t)\, \chi_{I_1}\dots\chi_{I_n}\right)_\1\,,
\label{eq:superpotentialGen}
\ee
%
where $\1$ stands for an invariant singlet of $\G_N$.
From eq.~\eqref{eq:modtransforms} it is clear that 
the invariance of $W$ requires the $Y_{I_1\,\dots\,I_n}(\t)$ to transform in the following way: 
\be
Y_{I_1\,\dots\,I_n}(\t) \rightarrow (c\t + d)^{k_{Y}} \rho_{Y}(\g)\, Y_{I_1\,\dots\,I_n}(\t)\,,
\ee
%
where $\rho_{Y}$ is a representation of $\G_N$, and $k_{Y}$ and $\rho_{Y}$ 
are such that
\begin{align}
&k_{Y} = k_{I_1} + \dots + k_{I_n}\,, \\[2mm]
&\rho_{Y} \ot \rho_{I_1} \ot \dots \ot \rho_{I_n} \supset \1\,.
\end{align}
%
Thus, $Y_{I_1\,\dots\,I_n}(\t)$ are modular forms of weight $k_{Y}$ and level $N$ 
furnishing the representation $\rho_{Y}$ of the finite modular group $\G_N$
(cf.~eq.~\eqref{eq:vvmodforms}).

\subsection{Modular forms of level 4}
\label{subsec:modforms4}
 The dimension of the linear space formed by 
the modular forms of weight 2 and level 4 is equal to 5
(see, e.g., \cite{Feruglio:2017spp}), i.e., there are five
linearly independent modular forms of the lowest non-trivial weight.
In Ref.~\cite{Penedo:2018nmg} these forms have been explicitly
constructed in terms of the Dedekind eta function 
\be
\eta(\t) \equiv q^{1/24} \prod_{n=1}^{\infty}\left(1 - q^n\right)\,,
\qquad
q = e^{2\pi i\t}\,.
\ee
%
Namely, defining
\begin{align}
Y(a_1,\dots,a_6|\t) &\equiv \frac{\di}{\di\t} \bigg[
a_1 \log \eta\left(\t+\frac{1}{2}\right) +
a_2 \log \eta\left(4\t\right) +
a_3 \log \eta\left(\frac{\t}{4}\right) \nonumber\\
&+ a_4 \log \eta\left(\frac{\t+1}{4}\right) + 
a_5 \log \eta\left(\frac{\t+2}{4}\right) +
a_6 \log \eta\left(\frac{\t+3}{4}\right)\bigg]\,,
\end{align}
%
with $a_1+\dots+a_6 = 0$, the basis of the modular forms of weight 2 reads
\begin{align}
Y_1(\tau) &\equiv Y(1,1,\omega,\omega^2,\omega,\omega^2|\tau)\,,\\
Y_2(\tau) &\equiv Y(1,1,\omega^2,\omega,\omega^2,\omega|\tau)\,,\\
Y_3(\tau) &\equiv Y(1,{-1},{-1},{-1},1,1|\tau)\,,\\
Y_4(\tau) &\equiv Y(1,-1,-\omega^2,-\omega,\omega^2,\omega|\tau)\,,\\
Y_5(\tau) &\equiv Y(1,-1,-\omega,-\omega^2,\omega,\omega^2|\tau)\,,
\end{align}
%
with $\o \equiv e^{2\pi i/3}$. 
Furthermore, as shown in \cite{Penedo:2018nmg}, 
the $Y_1(\t)$ and $Y_2(\t)$ form a doublet transforming in the $\2$ of $S_4$, 
while the three remaining modular forms
make up a triplet
transforming in $\tp$ of $S_4$.
In what follows, we denote the doublet and the triplet as 
\be
Y_\2(\tau) \equiv
\begin{pmatrix}Y_1(\tau)\\ Y_2(\tau)\end{pmatrix}\,,
\qquad
Y_\tp(\tau) \equiv
\begin{pmatrix}Y_3(\tau)\\ Y_4(\tau)\\ Y_5(\tau)\end{pmatrix}\,.
\ee
%

 The modular forms of higher weights $k = 4,\,6,\dots$, 
can be built from the $Y_i(\t)$, $i=1,\dots,5$. 
Thus, the $Y_i(\t)$ generate the ring of all modular forms of level 4 
\be
\mathcal{M}(\G(4)) = \bigoplus_{k = 0}^\infty \mathcal{M}_k(\G(4))\,.
\ee
%
The dimension of the linear space $\mathcal{M}_k(\G(4))$ of modular forms 
of weight $k$ is $2k+1$. 
The modular forms of higher weight transform according to certain irreps of $S_4$. 
For example, at weight 4 we have 9 independent modular forms, 
which arrange themselves in an invariant singlet, a doublet and two triplets 
transforming in the $\1$, $\2$, $\3$ and $\tp$ irreps of $S_4$, respectively~\cite{Penedo:2018nmg}:
\begin{equation}
\begin{aligned}
Y^{(4)}_\1 = Y_1 Y_2\,,&
\qquad
Y^{(4)}_\2 = \begin{pmatrix}
Y_2^2 \\
Y_1^2
\end{pmatrix}, \\
Y^{(4)}_\3 = \begin{pmatrix}
Y_1 Y_4 - Y_2 Y_5 \\
Y_1 Y_5 - Y_2 Y_3 \\
Y_1 Y_3 - Y_2 Y_4
\end{pmatrix},&
\qquad
Y^{(4)}_\tp = \begin{pmatrix}
Y_1 Y_4 + Y_2 Y_5 \\
Y_1 Y_5 + Y_2 Y_3 \\
Y_1 Y_3 + Y_2 Y_4
\end{pmatrix}.
\label{eq:weight4}
\end{aligned}
\end{equation}
%
Some higher weight multiplets are given in Appendix~\ref{app:higherweight}.
 In the next section we use the modular forms of level 4 to build
a modular-invariant superpotential, as in eq.~\eqref{eq:superpotentialGen}.

\section{Seesaw Models without Flavons}
\label{sec:seesaw}
 We assume that neutrino masses originate from 
the (supersymmetric) type I seesaw mechanism. 
In this case, the superpotential in the lepton sector reads 
\be
W = \a \left(\Ec\, L\, \Hd\,  f_E\left(Y\right)\right)_\1 
+ g \left(\Nc\, L\, \Hu\,  f_N\left(Y\right)\right)_\1 
+ \L \left(\Nc\, \Nc\, f_M\left(Y\right)\right)_\1\,,
\ee
where a sum over all independent invariant singlets 
with the coefficients $\a = (\a, \a^\prime,\dots)$, $g = (g, g^\prime,\dots)$ and $\L = (\L, \L^\prime,\dots)$ 
is implied.
Here, $f_{E,\,N,\,M}(Y)$ denote the modular form multiplets
required to ensure modular invariance.

 For the sake of simplicity, we will make the following assumptions: 
\begin{itemize}
\item Higgs doublets $\Hu$ and $\Hd$ transform trivially under $\G_4$, $\rho_u = \rho_d \sim \1$, 
and $k_u = k_d = 0$;
\item lepton $SU(2)$ doublets $L_1$, $L_2$, $L_3$ furnish a 3-dimensional irrep  of $\G_4$, i.e., $\rho_L \sim \3$ or $\tp$;
\item neutral lepton gauge singlets 
$\Nc_1$, $\Nc_2$, $\Nc_3$ transform as a triplet  of $\G_4$, 
$\rho_N \sim \3$ or $\tp$;
\item charged lepton $SU(2)$ singlets $\Ec_1$, $\Ec_2$, $\Ec_3$ transform as singlets of $\G_4$, 
$\rho_{1,2,3} \sim \1,\,\op$.
\end{itemize} 

 With these assumptions, we can rewrite the superpotential as 
\be
W = \sum_{i=1}^{3} \a_i \left(\Ec_i\, L\, f_{E_i}\left(Y\right)\right)_\1 \Hd 
+ g \left(\Nc\, L\, f_N\left(Y\right)\right)_\1  \Hu
+ \L \left(\Nc\, \Nc\, f_M\left(Y\right)\right)_\1\,,
\label{eq:superpotential}
\ee
where the sum over all independent singlet contributions is understood 
as specified earlier.
Assigning weights $(-k_i)$, $(-k_L)$, $(-k_N)$ to $\Ec_i$, $L$, $\Nc$, 
and weights $k_{\a_i}$, $k_g$, $k_\L$ to the multiplets of modular forms 
$f_{E_i}(Y)$, $f_N(Y)$, $f_M(Y)$, modular invariance of the superpotential requires
\be
\begin{cases}
k_{\a_i} = k_i + k_L \\
k_g = k_N + k_L \\
k_\L = 2\,k_N
\end{cases}
\Leftrightarrow\quad
\begin{cases}
k_i = k_{\a_i} - k_g + k_\L/2 \\
k_L = k_g - k_\L/2 \\
k_N = k_\L/2
\end{cases}\,.
\label{eq:weights}
\ee
Thus, by specifying the weights of the modular forms one obtains the weights 
of the matter superfields. 

 After modular symmetry breaking, the matrices 
of charged lepton and neutrino Yukawa couplings, $\l$ and $\y$,
as well as the Majorana mass matrix $M$ for heavy neutrinos, are generated:
\be
W = \l_{ij}\, \Ec_i\, L_j\, \Hd 
+ \y_{ij}\, \Nc_i\, L_j\, \Hu
+ \frac{1}{2}\, M_{ij}\, \Nc_i\, \Nc_j\,,
\ee
where a sum over $i,j = 1,2,3$ is assumed. 
Eventually, after integrating out $\Nc$ and after electroweak symmetry breaking,
the charged lepton mass matrix $M_e$ and the light neutrino 
Majorana mass matrix $M_\nu$ are generated:%
\footnote{We work in the left-right convention for the charged 
lepton mass term and 
the right-left convention for the light and heavy neutrino Majorana mass terms.}
\begin{align}
M_e &= \vd\, \l^\dagger\,, \\
M_\nu &= - \vu^2\, \y^T M^{-1} \y\,,
\label{eq:MnuSeesaw}
\end{align}
with $\vd \equiv \vev{\Hd^0}$ and $\vu \equiv \vev{\Hu^0}$.
In what follows we will systematically consider low weights 
$k_{\a_i}$, $k_g$, $k_\L$ and identify the corresponding seesaw models.

\subsection{The Majorana mass term for heavy neutrinos}
 We start with the analysis of 
the Majorana mass term for heavy neutrinos. 
If $k_\L = 0$, i.e., no non-trivial modular forms are present 
in the last term of eq.~\eqref{eq:superpotential}, 
$k_N = 0$, and for both choices $\rho_N \sim \3$ or $\rho_N \sim \tp$ we have 
\be
\left(\Nc\, \Nc\right)_\1 = \Nc_1\, \Nc_1 + \Nc_2\, \Nc_3 + \Nc_3\, \Nc_2\,,
\label{eq:3per3}
\ee
which leads to the following mass matrix for heavy neutrinos:
\be
M = 2\,\L \begin{pmatrix}
1 & 0 & 0 \\
0 & 0 & 1 \\
0 & 1 & 0
\end{pmatrix},~~\text{for}~~k_\L=0\,.
\label{eq:MkL0}
\ee
%
Thus, in this case, the spectrum of heavy neutrino masses is degenerate, 
and the only free parameter is the overall scale $\L$, which can be
rendered real. 
The Majorana mass term with the mass matrix in eq.~(\ref{eq:MkL0}) 
conserves a ``non-standard'' lepton charge
and two of the three heavy Majorana neutrinos with definite mass
form a Dirac pair \cite{Leung:1983ti}.

 Allowing for modular forms of weight $k_\L = 2$ in the Majorana mass term, 
we have instead the following structure in the superpotential: 
\be
\L \left(\Nc\, \Nc\, Y_\2\right)_\1 + \L^\prime \left(\Nc\, \Nc\, Y_\tp\right)_\1\,.
\ee
The second term vanishes because the $\tp$ from the decomposition of 
$\3\ot\3$ ($\tp \ot \tp$) needed to form an invariant singlet is antisymmetric 
(see Appendix~\ref{app:CGcoefficients}).
Applying the decomposition rules to the first term, we obtain
\be
M = 2\, \L \begin{pmatrix}
0 & Y_1 & Y_2 \\
Y_1 & Y_2 & 0 \\
Y_2 & 0 & Y_1
\end{pmatrix},~~\text{for}~~k_\L=2\,,
\label{eq:MkL2}
\ee
%
where $Y_{1,2}$ depend on the complex VEV of $\tau$.
Therefore, there are 3 free real parameters in the matrix $M$.

Increasing $k_\L$ to $4$ leads to a bigger number of free parameters, since 
more than one invariant singlet can be formed. 
There are nine independent modular forms of weight 4 and level 4.
As shown in \cite{Penedo:2018nmg}, they are organised in an invariant singlet, 
a doublet and two triplets, one transforming in the $\3$ and the other in the $\tp$ of $\G_4$, cf.~eq.~\eqref{eq:weight4}.
Hence, the relevant part of $W$ reads
\be
\L \left(\Nc\, \Nc\, Y^{(4)}_\1\right)_\1 + \Lp \left(\Nc\, \Nc\, Y^{(4)}_\2\right)_\1 
+ \Lpp \left(\Nc\, \Nc\, Y^{(4)}_\3\right)_\1 
+ \Lppp \left(\Nc\, \Nc\, Y^{(4)}_\tp\right)_\1\,.
\ee
The last term vanishes, as before, due to antisymmetry.
The remaining three terms lead to 
\begin{align}
M &= 2\,\L \left[
Y_1 Y_2 \begin{pmatrix}
1 & 0 & 0 \\
0 & 0 & 1 \\
0 & 1 & 0
\end{pmatrix} +
\frac{\Lp}{\L}
\begin{pmatrix}
0 & Y_2^2 & Y_1^2 \\
Y_2^2 & Y_1^2 & 0 \\
Y_1^2 & 0 & Y_2^2
\end{pmatrix} \right. \nonumber\\
& + \frac{\Lpp}{\L} \left. \begin{pmatrix}
2 \left(Y_1Y_4 - Y_2Y_5\right) & Y_2Y_4 - Y_1Y_3 & Y_2Y_3 - Y_1Y_5 \\ 
Y_2Y_4 - Y_1Y_3 & 2 \left(Y_1Y_5 - Y_2Y_3\right) & Y_2Y_5 - Y_1Y_4 \\ 
Y_2Y_3 - Y_1Y_5 & Y_2Y_5 - Y_1Y_4 & 2\left(Y_1Y_3 - Y_2Y_4\right)
\end{pmatrix}
\right],~~\text{for}~~k_\L=4\,.
\label{eq:MkL4}
\end{align}
Thus, apart from $\vev{\tau}$, there are one real ($\L$) and two complex 
($\Lp/\L$, $\Lpp/\L$) free parameters in $M$,
that is, 5 real parameters apart from $\vev{\tau}$.
Weight 6 and higher weight modular forms
(see Appendix~\ref{app:higherweight})
will lead to more free parameters and thus to a decrease in predictivity.

\subsection{The neutrino Yukawa couplings}
 Next we analyse the neutrino Yukawa interaction term 
 in the superpotential of eq.~\eqref{eq:superpotential}. 
If $k_g = 0$, the irreps in which $\Nc$ and $L$ transform 
should be the same to construct an invariant singlet, i.e., $\rho_N = \rho_L \sim \3$ or $\tp$. 
The structure of the singlet is the same of eq.~\eqref{eq:3per3}, 
and the neutrino Yukawa matrix reads
\be
\y = g \begin{pmatrix}
1 & 0 & 0 \\
0 & 0 & 1 \\
0 & 1 & 0
\end{pmatrix},~~\text{for}~~k_g=0\,.
\label{eq:ykg0}
\ee
 
 The lowest non-trivial weight, $k_g = 2$, leads to
\be
g \left(\Nc\, L\, Y_\2\right)_\1 \Hu + \gp \left(\Nc\, L\, Y_\tp\right)_\1 \Hu\,.
\ee
There are 4 possible assignments of $\rho_N$ and $\rho_L$ we consider. 
Two of them, namely $\rho_N = \rho_L \sim \3$ and $\rho_N = \rho_L \sim \tp$ 
give the following form of $\y$:
\be
\y = g \left[  
\begin{pmatrix}
0 & Y_1 & Y_2 \\
Y_1 & Y_2 & 0 \\
Y_2 & 0 & Y_1
\end{pmatrix} 
+ \frac{\gp}{g} \begin{pmatrix}
0 & Y_5 & -Y_4 \\
- Y_5 & 0 & Y_3 \\
Y_4 & -Y_3 & 0
\end{pmatrix}
\right],~~\text{for}~~k_g=2~~\text{and}~~\rho_N = \rho_L\,.
\label{eq:ykg2eq}
\ee
The two remaining combinations, $(\rho_N,\rho_L) \sim (\3,\tp)$ and $(\tp,\3)$, lead to:
\be
\y = g \left[ 
\begin{pmatrix}
0 & -Y_1 & Y_2 \\
-Y_1 & Y_2 & 0 \\
Y_2 & 0 & -Y_1
\end{pmatrix} 
+ \frac{\gp}{g} \begin{pmatrix}
2Y_3 & -Y_5 & -Y_4 \\
-Y_5 & 2Y_4 & -Y_3 \\
-Y_4 & -Y_3 & 2Y_5
\end{pmatrix}
\right],~~\text{for}~~k_g=2~~\text{and}~~\rho_N \neq \rho_L\,.
\label{eq:ykg2neq}
\ee
In both cases, up to an overall factor, the matrix $\y$ depends on one 
complex parameter $\gp/g$ and the VEV $\vev{\tau}$.

 Considering further the case of $k_g = 4$, we have
\be
\left[g \left(\Nc\, L\, Y^{(4)}_\1\right)_\1 
+ \gp \left(\Nc\, L\, Y^{(4)}_\2\right)_\1 
+ \gpp \left(\Nc\, L\, Y^{(4)}_\3\right)_\1 
+ \gppp \left(\Nc\, L\, Y^{(4)}_\tp\right)_\1
\right] \Hu\,.
\label{eq:kg4}
\ee
Again we have two equivalent possibilities with $\rho_N = \rho_L$ 
and two others with $\rho_N \neq \rho_L$.
In the former case, the matrix $\y$ reads
\begin{align}
\y &= g \left[
Y_1 Y_2 \begin{pmatrix}
1 & 0 & 0 \\
0 & 0 & 1 \\
0 & 1 & 0
\end{pmatrix} +
\frac{g'}{g}
\begin{pmatrix}
0 & Y_2^2 & Y_1^2 \\
Y_2^2 & Y_1^2 & 0 \\
Y_1^2 & 0 & Y_2^2
\end{pmatrix} \right. \nonumber\\
& + \frac{\gpp}{g} \begin{pmatrix}
2 \left(Y_1Y_4 - Y_2Y_5\right) & Y_2Y_4 - Y_1Y_3 & Y_2Y_3 - Y_1Y_5 \\ 
Y_2Y_4 - Y_1Y_3 & 2 \left(Y_1Y_5 - Y_2Y_3\right) & Y_2Y_5 - Y_1Y_4 \\ 
Y_2Y_3 - Y_1Y_5 & Y_2Y_5 - Y_1Y_4 & 2\left(Y_1Y_3 - Y_2Y_4\right)
\end{pmatrix} \nonumber\\
& + \frac{\gppp}{g} \left. \begin{pmatrix}
0 & Y_1Y_3 + Y_2Y_4 & - Y_1Y_5 - Y_2Y_3 \\ 
- Y_1Y_3 - Y_2Y_4 & 0 & Y_1Y_4 + Y_2Y_5 \\ 
Y_1Y_5 + Y_2Y_3 & - Y_1Y_4 - Y_2Y_5 & 0
\end{pmatrix} \right],~~\text{for}~~k_g=4~~\text{and}~~\rho_N = \rho_L\,.
\end{align}
It depends on 7 real parameters and the complex $\vev{\tau}$.
In the case of different representations $\rho_N \neq \rho_L$, 
$\3 \ot \tp$ does not contain the invariant singlet, 
such that the first term in eq.~\eqref{eq:kg4} is not possible. 
The sum of three remaining terms yields
\begin{align}
\y &= \gp \left[\begin{pmatrix}
0 & - Y_2^2 & Y_1^2 \\
- Y_2^2 & Y_1^2 & 0 \\
Y_1^2 & 0 & - Y_2^2
\end{pmatrix} \right. 
\pm \frac{\gpp}{\gp} \begin{pmatrix}
0 & Y_1Y_3 - Y_2Y_4 & Y_2Y_3 - Y_1Y_5 \\ 
Y_2Y_4 - Y_1Y_3 & 0 & Y_1Y_4 - Y_2Y_5 \\ 
Y_1Y_5 - Y_2Y_3 & Y_2Y_5 - Y_1Y_4 & 0
\end{pmatrix} \nonumber\\
& + \frac{\gppp}{\gp} \left. \begin{pmatrix}
2 \left(Y_1Y_4 + Y_2Y_5\right) & - Y_1Y_3 - Y_2Y_4 & - Y_1Y_5 - Y_2Y_3 \\ 
- Y_1Y_3 - Y_2Y_4 & 2 \left(Y_1Y_5 + Y_2Y_3\right) & - Y_1Y_4 - Y_2Y_5 \\ 
- Y_1Y_5 - Y_2Y_3 & - Y_1Y_4 - Y_2Y_5 & 2\left(Y_1Y_3 + Y_2Y_4\right)
\end{pmatrix} \right],~\text{for}~k_g=4~\text{and}~\rho_N \neq \rho_L\,,
\end{align}
where plus sign in $\pm$ corresponds to $(\rho_N,\rho_L) \sim (\3,\tp)$ 
and minus sign to $(\rho_N,\rho_L) \sim (\tp,\3)$. 
This minus sign can be absorbed in $\gpp$. 
Thus, apart from $\vev{\tau}$, the matrix $\y$ depends on 5 real parameters. 
Given the rising multiplicity of free parameters, we do not consider weights $k_g$ higher than 4 in the present analysis.

\subsection{The charged lepton Yukawa couplings}
 Further we investigate the charged lepton Yukawa interaction 
terms in the superpotential.
Since we consider $\rho_i \sim \1$ or $\op$ and $\rho_L \sim \3$ or $\tp$, 
we have four possible combinations $\rho_i \ot \rho_L$. None of them contain 
the invariant singlet. Thus, the weights $k_{\a_i}$ cannot be zero, i.e., 
they are strictly positive, $k_{\a_i} > 0$. 
Moreover, $f_{E_i}\left(Y\right)$ should transform in $\3$ 
if $(\rho_i,\rho_L) \sim (\1,\3)$ or $(\op,\tp)$, 
and in $\tp$ if $(\rho_i,\rho_L) \sim (\1,\tp)$ or $(\op,\3)$.
Thus, for each $i=1,2,3$, we have
\be
\a_i \left(\Ec_i\, L\, f_{E_i}\left(Y\right)\right)_\1 \Hd = 
\Ec_i\, \sum_{a} \a_{i,a} \left[
L_1 \left(Y_a^{(k_{\a_i})}\right)_1 +
L_2 \left(Y_a^{(k_{\a_i})}\right)_3 + 
L_3 \left(Y_a^{(k_{\a_i})}\right)_2
\right] \Hd\,,
\label{eq:CLinvariants}
\ee
where $Y_a^{(k_{\a_i})}$ are independent triplets 
($\3$ or $\tp$ depending on $\rho_i$ and $\rho_L$) of weight $k_{\a_i}$. 

 There exists only one triplet 
$(Y_3, Y_4, Y_5)^T \sim \tp$ of the lowest non-trivial weight 2. 
If $k_{\a_i} = 2$, eq.~\eqref{eq:CLinvariants} 
reads
\be
\a_i \left(\Ec_i\, L\, Y_\tp\right)_\1 \Hd = 
\a_i\, \Ec_i\, \left[
L_1\, Y_3 + L_2\, Y_5 + L_3\, Y_4
\right] \Hd\,.
\ee
Therefore, if $k_{\a_i} = 2$ for all $i=1,2,3$, three rows of the charged lepton Yukawa matrix $\l$ 
will be proportional to each other, and $\rank(\l) = 1$ 
implying that two of the three charged lepton masses are zero,
since $\rank(\l) = \rank(\l^\dagger\l)$. 
If $k_{\a_i} = k_{\a_j} = 2$, where $i \neq j$, and $k_{\a_p} > 2$,
one has $\rank(\l) = 2$, i.e., one of the masses is zero. 
Thus, in order to have maximal rank, $\rank(\l) = 3$, and no zero masses, 
only one $k_{\a_i}$ can be equal to 2. 

 The minimal (in terms of weights) possibility 
is defined by $k_{\a_i} = 2$ and $k_{\a_j} = k_{\a_p} = 4$, for $j \neq p$. 
Indeed, there are two triplets of weight 4, namely $Y^{(4)}_\3$ and $Y^{(4)}_\tp$.
To avoid having a reduced $\rank(\l)$, the representations 
$\rho_j$ and $\rho_p$ should be different. 
This ensures that both $Y^{(4)}_\3$ and $Y^{(4)}_\tp$ are present 
in the superpotential, and the corresponding rows in the matrix $\l$ 
are linearly independent. 
Then the relevant part of $W$, which we denote as $W_e$, 
takes one of the following 6 forms:
\begin{align}
\a \left(\Ec_1\, L\, Y_\tp\right)_\1 \Hd + 
\b \left(\Ec_2\, L\, Y^{(4)}_\3\right)_\1 \Hd +
\g \left(\Ec_3\, L\, Y^{(4)}_\tp\right)_\1 \Hd\,, 
\label{eq:CLform1}\\
\a \left(\Ec_1\, L\, Y_\tp\right)_\1 \Hd + 
\b \left(\Ec_2\, L\, Y^{(4)}_\tp\right)_\1 \Hd +
\g \left(\Ec_3\, L\, Y^{(4)}_\3\right)_\1 \Hd\,, 
\label{eq:CLform2}\\
\a \left(\Ec_1\, L\, Y^{(4)}_\3\right)_\1 \Hd + 
\b \left(\Ec_2\, L\, Y_\tp\right)_\1 \Hd +
\g \left(\Ec_3\, L\, Y^{(4)}_\tp\right)_\1 \Hd\,, 
\label{eq:CLform3}\\
\a \left(\Ec_1\, L\, Y^{(4)}_\tp\right)_\1 \Hd + 
\b \left(\Ec_2\, L\, Y_\tp\right)_\1 \Hd +
\g \left(\Ec_3\, L\, Y^{(4)}_\3\right)_\1 \Hd\,, 
\label{eq:CLform4}\\
\a \left(\Ec_1\, L\, Y^{(4)}_\3\right)_\1 \Hd + 
\b \left(\Ec_2\, L\, Y^{(4)}_\tp\right)_\1 \Hd +
\g \left(\Ec_3\, L\, Y_\tp\right)_\1 \Hd\,, 
\label{eq:CLform5}\\
\a \left(\Ec_1\, L\, Y^{(4)}_\tp\right)_\1 \Hd + 
\b \left(\Ec_2\, L\, Y^{(4)}_\3\right)_\1 \Hd +
\g \left(\Ec_3\, L\, Y_\tp\right)_\1 \Hd\,. 
\label{eq:CLform6}
\end{align}
The possible assignments of irreps to the $L$ and $\Ec_i$ superfields for each of these forms of $W_e$ are given in Table~\ref{tab:CLirreps}.
\begin{table}
\centering
\begin{tabular}{lllll}
\toprule
$W_e$ & $\rho_L$ & $\rho_1$ & $\rho_2$ & $\rho_3$ \\ 
\midrule 
\multirow{2}{*}{eqs.~\eqref{eq:CLform1}, \eqref{eq:CLform6}} & 
$\3$ & $\op$ & $\1$ & $\op$ \\ 
 & $\tp$ & $\1$ & $\op$ & $\1$ \\
\midrule
\multirow{2}{*}{eqs.~\eqref{eq:CLform2}, \eqref{eq:CLform4}} & 
$\3$ & $\op$ & $\op$ & $\1$ \\ 
 & $\tp$ & $\1$ & $\1$ & $\op$ \\
\midrule
\multirow{2}{*}{eqs.~\eqref{eq:CLform3}, \eqref{eq:CLform5}} & 
$\3$ & $\1$ & $\op$ & $\op$ \\ 
 & $\tp$ & $\op$ & $\1$ & $\1$ \\
\bottomrule
\end{tabular}
\caption{The possible assignments of irreps 
for the $L$, $\Ec_1$, $\Ec_2$ and $\Ec_3$ superfields
in the described minimal set-up. 
For each form of $W_e$, the upper and lower lines lead to 
the same results for the matrix $\l$.}
\label{tab:CLirreps}
\end{table}
%
Equation~\eqref{eq:CLform1} leads to 
\be
\l = \begin{pmatrix}
\a\, Y_3 & \a\, Y_5 & \a\, Y_4 \\
\b \left(Y_1Y_4 - Y_2Y_5\right) & \b \left(Y_1Y_3 - Y_2Y_4\right) & \b \left(Y_1Y_5 - Y_2Y_3\right) \\
\g \left(Y_1Y_4 + Y_2Y_5\right) & \g \left(Y_1Y_3 + Y_2Y_4\right) & \g \left(Y_1Y_5 + Y_2Y_3\right) \\
\end{pmatrix}\,,
\label{eq:lambda1}
\ee
while the other 5 forms of $W_e$ yield a $\l$ which differs 
from that in eq.~\eqref{eq:lambda1} 
by permutations of the rows (and renaming of the free parameters). 
However, those permutations do not affect 
the matrix $U_e$ diagonalising $M_e M_e^\dagger = \vd^2\, \l^\dagger \l$, 
and thus do not lead to new results for the PMNS matrix.
In what follows, 
without loss of generality,
we adhere to the minimal choice in eq.~\eqref{eq:CLform1}, taking $k_{\a_1} = 2$ and $k_{\a_2} = k_{\a_3} = 4$. 
As we can see, in this ``minimal'' example the matrix $\l$ depends 
on 3 free parameters, $\a$, $\b$ and $\g$, 
which can be rendered real by re-phasing of the charged lepton fields, 
and the complex $\vev{\tau}$. 

 The next natural choice of the weights would be $k_{\a_i} = 4$ for any $i=1,2,3$. 
However, such a combination of weights leads to $\rank(\l) < 3$, 
since there are only two independent triplets 
$Y^{(4)}_\3$ and $Y^{(4)}_\tp$ of weight 4. 
Hence, for further choices of the $k_{\a_i}$ at least one of them should equal 6.

\subsection{Summary of models}
 Let us bring together the different pieces we have obtained so far 
and summarise the number of free and independent real parameters 
in the models containing modular forms of weights $\leq 4$.
Apart from the dependence of $M_\nu$ and $M_e$ on $\vev{\tau}$
(2 real parameters),%
\footnote{In the case $k_\L = k_g = 0$, $M_\nu$ does not depend on $\vev{\tau}$.}
we have 3 real parameters $\a$, $\b$ and $\g$ from 
the charged lepton sector. 
Making use of eq.~\eqref{eq:MnuSeesaw}, we can count 
the number of free parameters in the light neutrino Majorana mass matrix $M_\nu$
for different combinations of $k_\L$ and $k_g$. 
We present the results in Table~\ref{tab:freeparams}.
\begin{table}
\centering
\renewcommand{\arraystretch}{1.2}
\begin{tabular}{l|lll}
\toprule
\diagbox{$k_g\phantom{99}$}{$k_\L\phantom{99}$} & $0$ & $2$ & $4$ \\ 
\midrule 
$0$ & $1~(1)~[6]$ & $1~(3)~[6]$  & $5~(7)~[10]$ \\ 
$2$ & $3~(5)~[8]$ & $3~(5)~[8]$  & $7~(9)~[12]$ \\ 
$4$, $\rho_N \neq \rho_L$ & $5~(7)~[10]$ & $5~(7)~[10]$ & $9~(11)~[14]$ \\ 
$4$, $\rho_N = \rho_L$ & $7~(9)~[12]$ & $7~(9)~[12]$ & $11~(13)~[16]$ \\ 
\bottomrule
\end{tabular}
\caption{Number of free independent real parameters in models containing 
modular forms of weights $\leq 4$.
For each pair $(k_\L,k_g)$, the first number
is the number of parameters in $M_\nu$ apart from  $\vev{\tau}$. 
The second number (in parentheses) is 
the number of parameters in $M_\nu$ including the 2 real parameters from $\vev{\tau}$.
The third number [in brackets] is 
the total number of free independent parameters contained in $M_\nu$ and $M_e$.}
\label{tab:freeparams}
\end{table}

 In the case of $k_\L = k_g = 0$,  
the light neutrino mass matrix has the following form 
(see eqs.~\eqref{eq:MnuSeesaw}, \eqref{eq:MkL0} and \eqref{eq:ykg0}): 
\be
M_\nu = - \frac{g^2 v_u^2}{2\,\L} 
\begin{pmatrix}
1 & 0 & 0 \\
0 & 0 & 1 \\
0 & 1 & 0
\end{pmatrix},
\ee
%
which leads to $|m_1| = |m_2| = |m_3|$ 
in contradiction with the neutrino oscillation data. 
In the cases of $k_\L = k_g = 4$, the total number of 
free independent real parameters is bigger than 12, 
i.e., than the number of observables we want to describe or predict. 
The observables are 3 charged lepton masses, 3 neutrino masses, 
and 3 mixing angles, 1 Dirac and 2 Majorana~\cite{Bilenky:1980cx} CPV phases 
in the PMNS matrix. 
In the next section we will investigate in detail
potentially viable models
with both $k_\L$ and $k_g \leq 2$.

\section{Numerical Analysis}
\label{sec:numerical}
 Each of the investigated models depends on a set of dimensionless parameters
\begin{equation}
  p_i = (\tau,\, \beta/\alpha,\, \gamma/\alpha,\, g'/g,\, \ldots,\, \Lambda'/\Lambda,\, \ldots)\,,
  \label{eq:params}
\end{equation}
which determine dimensionless observables (mass ratios, mixing angles and phases), and two overall mass scales: $v_d\, \alpha$ for $M_e$ and $v_u^2\, g^2 / \Lambda$ for $M_{\nu}$.
Phenomenologically viable models are those that lead to values of observables 
which are in close agreement 
with the experimental results summarised in
Table~\ref{tab:globalFit}.%
\footnote{The atmospheric mass-squared difference 
$\Delta m^2_{31} = \Delta m^2 + \delta m^2/2$ 
for the NO spectrum of light neutrino masses and 
$\Delta m^2_{32} = \Delta m^2 - \delta m^2/2$ for the IO spectrum.
We assume also to be in a regime in which the running of neutrino parameters is negligible (see Section~\ref{sec:corrections} for a discussion of renormalisation group corrections).
}
\begin{table}
\centering
\renewcommand{\arraystretch}{1.2}
\begin{tabular}{l|cc} 
\toprule
Observable & \multicolumn{2}{c}{Best fit value and $1\sigma$ range} \\ 
\midrule
$m_e / m_\mu$ & \multicolumn{2}{c}{$0.0048 \pm 0.0002$} \\
$m_\mu / m_\tau$ & \multicolumn{2}{c}{$0.0565 \pm 0.0045$} \\ 
\midrule
& NO & IO \\
$\delta m^2/(10^{-5}\text{ eV}^2)$ & \multicolumn{2}{c}{$7.34^{+0.17}_{-0.14}$} \\
$|\Delta m^2|/(10^{-3}\text{ eV}^2)$ & $2.455^{+0.035}_{-0.032}$ & $2.441^{+0.033}_{-0.035}$ \\
$r \equiv \delta m^2/|\Delta m^2|$ & $0.0299\pm0.0008$ & $0.0301\pm0.0008$\\
$\sin^2\theta_{12}$ & $0.304^{+0.014}_{-0.013}$ & $0.303^{+0.014}_{-0.013}$ \\
$\sin^2\theta_{13}$ & $0.0214^{+0.0009}_{-0.0007}$ & $0.0218^{+0.0008}_{-0.0007}$ \\
$\sin^2\theta_{23}$ & $0.551^{+0.019}_{-0.070}$ & $0.557^{+0.017}_{-0.024}$ \\
$\delta/\pi$ & $1.32^{+0.23}_{-0.18}$  & $1.52^{+0.14}_{-0.15}$ \\
\bottomrule
\end{tabular}
\caption{Best fit values and 1$\sigma$ ranges for 
neutrino oscillation parameters, obtained from the global analysis
of Ref.~\cite{Capozzi:2018ubv}, and for charged lepton mass ratios,
given at the scale $2\times 10^{16}$ GeV with the $\tan \beta$ averaging
described in~\cite{Feruglio:2017spp}, obtained from Ref.~\cite{Ross:2007az}.
The parameters entering the definition of $r$ are $\delta m^2 \equiv m_2^2-m_1^2$
and $\Delta m^2 \equiv m_3^2 - (m_1^2+m_2^2)/2$.
The best fit value and $1\sigma$ range of $\delta$
did not drive the numerical searches here reported.}
\label{tab:globalFit}
\end{table}
%

As a measure of goodness of fit, we use the sum of one-dimensional $\Delta \chi^2_j$ functions 
\be
  \Delta \chi^2(p_i) = \sum_{j=1}^6 \Delta \chi_j^2(p_i)\,,
  \label{eq:chisq}
\ee
for six accurately known dimensionless
\footnote{If a model successfully reproduces dimensionless observables, the overall mass scales can be easily recovered by fitting them to the charged lepton masses \(m_e,\, m_{\mu},\, m_{\tau},\) and the neutrino 
mass-squared differences \(\delta m^2\) and \(|\Delta m^2|\).} observable quantities
\be
  q_j = (m_e/m_{\mu},\, m_{\mu}/m_{\tau},\, r,\, \sin^2 \theta_{12},\, \sin^2 \theta_{13}, \sin^2 \theta_{23})\,.
\ee
In eq.~\eqref{eq:chisq} 
we have assumed approximate independence 
of the fitted quantities (observables).
In what follows, we define \(N\sigma \equiv \sqrt{\Delta \chi^2}\).
For \(\sin^2\theta_{ij}\), we make use of the one-dimensional projections $\Delta {\chi}_j^2$ ($j=4,5,6$) from Ref.~\cite{Capozzi:2018ubv},%
\footnote{These one-dimensional 
$\Delta\chi^2_j$ ($j=4,5,6$) projections
were kindly shared with us by 
the authors of Ref.~\cite{Capozzi:2018ubv}, 
and they are represented in Fig.~3 
of this reference.}
whereas for the remaining quantities we employ the Gaussian approximation:
\begin{equation}
  \Delta {\chi}_j^2(p_i) = \left( \frac{q_j(p_i) - q_{j, \text{best fit}}}{\sigma_j} \right)^2\,,
  \quad j = 1,2,3\,.
\end{equation}

We restrict the parameter space in the following way:
\be
\begin{aligned}
&\log_{10} (\beta/\alpha)\,, \quad
\log_{10} (\gamma/\alpha)\,, \quad
\log_{10} \left| g'/g \right|, \quad
\log_{10} \left| \Lambda'/\Lambda \right|, \ldots \in [-4, 4]\,,\\[2mm]
&\arg (g'/g)\,, \quad
\arg (\Lambda'/\Lambda)\,, \ldots \in [-\pi, \pi],
\end{aligned}
\label{eq:restriction}
\ee
and \(\tau\) is taken from the fundamental domain $\mathcal{D}$ of \(\overline{\Gamma}\),
\begin{equation}
  \label{eq:fundDomain}
  \mathcal{D} = \left\{ \tau \in \mathbb{C} :~~\im \tau > 0\,,~~|\re \tau| \leq \frac{1}{2}\,,~~|\tau| \geq 1 \right\},
\end{equation}
depicted in Fig.~\ref{fig:tau8param}, with an additional constraint of \(\im \tau \leq 2\).
The probability distribution for the numerical scan is chosen to be uniform with respect to the parameters in eq.~\eqref{eq:restriction} and to \(\re \tau\) and \(\im \tau\).
For further details of our numerical approach, see Appendix~\ref{app:numerical}.

Let us comment on why it is sufficient to scan $\tau$ in the fundamental domain~\eqref{eq:fundDomain}.
Since the underlying theory enjoys the modular symmetry $\overline{\Gamma}$, all the vacua related by modular transformations are physically equivalent.
Therefore, given a non-zero VEV of the modulus $\tau$, we can send it to $\tau' \in \mathcal{D}$ with a modular transformation.
This is similar to the choice of the Higgs doublet VEV in the Standard Model, which we can bring to its second component and make real by acting with a global gauge transformation.
Note however that couplings ($\alpha$, $\beta$, etc.) also transform non-trivially: the kinetic terms of the chiral supermultiplets arising from the Kähler potential in eq.~\eqref{eq:Kahler} should be rescaled to their canonical forms, and we implicitly absorb these rescalings into the couplings.
Since the kinetic term scalings change under modular transformations, one has to rescale the couplings accordingly, i.e.
\begin{equation}
  \tau \rightarrow \frac{a\tau + b}{c\tau + d} \quad \Rightarrow \quad g_i \to (c\tau+d)^{-k_{Y_i}} g_i\,,
\end{equation}
where $k_{Y_i}$ is the weight of the modular form corresponding to the coupling $g_i$.
For the models under investigation it means that dimensionless parameters in eq.~\eqref{eq:params} transform as
\begin{equation}
\begin{gathered}
  \left(\tau,\, \beta/\alpha,\, \gamma/\alpha,\, g'/g,\, \ldots,\, \Lambda'/\Lambda,\, \ldots\right) \to \\
  \left(\frac{a\tau + b}{c\tau + d}\,,\, (c\tau + d)^{-2}\, \beta/\alpha,\, (c\tau + d)^{-2}\, \gamma/\alpha,\, g'/g,\, \ldots,\, \Lambda'/\Lambda,\, \ldots\right).
\end{gathered}
\label{eq:equivParams}
\end{equation}
One can check that these two sets of parameters are physically equivalent, i.e., they lead to 
the same values of observables.

Another useful relation between different sets of parameters is a conjugation transformation 
defined as follows:
\begin{equation}
  \label{eq:conjParams}
  \begin{gathered}
    \left(\tau,\, \beta/\alpha,\, \gamma/\alpha,\, g'/g,\, \ldots,\, \Lambda'/\Lambda,\, \ldots\right) \to 
    \left(-\tau^{\star},\, \beta/\alpha,\, \gamma/\alpha,\, (g'/g)^{\star},\, \ldots,\, (\Lambda'/\Lambda)^{\star},\, \ldots\right)\,.
  \end{gathered}
\end{equation}
This transformation leaves all observables 
unchanged, except for the CPV phases, which flip their signs.
Therefore all the points we find in the following analysis come in pairs with the opposite CPV phases.

To see this, let us first notice that under $\tau \to -\tau^{\star}$ modular multiplets of weight 2 transform as
\begin{equation}
  Y_{\2,\tp}(\tau) \,\to\,
  Y_{\2,\tp}(-\tau^\star) =
  \left[ -\rho_{\2,\tp}\left(T^{-1}\right)\, Y_{\2,\tp}(\tau) \right]^{\star}
\end{equation}
(see Appendix~\ref{app:conjModularForms}), which is equivalent to:
\begin{enumerate}
\item a modular transformation $T^{-1}$,
\item change of sign $Y \to - Y$,
\item complex conjugation of the result.
\end{enumerate}
The first operation does not affect the physics as discussed earlier.
The effect of the second transformation can be absorbed into the unphysical phases for the mass matrices under consideration. 
Therefore $\tau \to -\tau^{\star}$ acts as complex conjugation on the modular forms.
Together with complex conjugation of couplings, it is nothing but complex conjugation of the mass matrices, which flips the signs of the CPV phases.
Inside the fundamental domain,
each viable $\vev{\tau}$ will thus be paired to $-\vev{\tau}^\star$, its
reflection across the imaginary axis.

\subsection{Models with 
\texorpdfstring{$(k_\Lambda, k_g) = (2,0)$}{(kΛ,kg)=(2,0)}}
In this case, the matrices are given by eqs.~\eqref{eq:MkL2}, \eqref{eq:ykg0} and \eqref{eq:lambda1}.
According to our numerical search, this model is unable to reproduce known data. The best point we have found is excluded at around 9.7 sigma confidence level, 
as it does not provide acceptable values of $\sin^2 \theta_{12}$ and $\sin^2 \theta_{23}$ (see Table~\ref{tab:case20predictions}).
\begin{table}
\centering
\renewcommand{\arraystretch}{1.2}
\begin{tabular}{c|c}
  \toprule
  $\re \tau$ & $\pm 0.4962$ \\
  $\im \tau$ & 1.208 \\
  $\beta/\alpha$ & 0.0002365 \\
  $\gamma/\alpha$ & 0.03178 \\
  $v_d\, \alpha$ [MeV] & 1059 \\
  $v_u^2\, g^2 / \Lambda$ [eV] & 0.1594 \\
  \midrule
  $m_e/m_{\mu}$ & 0.0048 \\
  $m_{\mu} / m_{\tau}$ & 0.0562 \\
  $r$ & 0.03003 \\
  $\delta m^2$ [$10^{-5} \text{ eV}^2$] & 7.334 \\
  $|\Delta m^2|$ [$10^{-3} \text{ eV}^2$] & 2.442 \\
  $\sin^2 \theta_{12}$ & {\color{red}0.5032} \\
  $\sin^2 \theta_{13}$ & 0.02235 \\
  $\sin^2 \theta_{23}$ & {\color{red}0.4021} \\
  \midrule
  Ordering & IO \\
  $m_1$ [eV] & 0.05981 \\
  $m_2$ [eV] & 0.06042 \\
  $m_3$ [eV] & 0.03423 \\
  $\textstyle \sum_i m_i$ [eV] & 0.1545 \\
  $|\langle m\rangle|$ [eV] & 0.04987 \\
  $\delta/\pi$ & $\pm 1.503$ \\
  $\alpha_{21}/\pi$ & $\pm 1.661$ \\
  $\alpha_{31}/\pi$ & $\pm 1.825$ \\
  \midrule
  $N \sigma$ & 9.657 \\
  \bottomrule
\end{tabular}
\caption{Best fit values of the parameters and observables 
in the models with \((k_{\Lambda}, k_g) = (2, 0)\). 
Here and in the following tables the weights $(k_{\a_1}, k_{\a_2}, k_{\a_3}) = (2,4,4)$.} 
\label{tab:case20predictions}
\end{table}
%

\subsection{Models with 
\texorpdfstring{$(k_\Lambda, k_g) = (0,2)$}{(kΛ,kg)=(0,2)}}
\begin{figure}[!ht]
\vskip -2.5cm
\centering
\includegraphics[width=\textwidth]{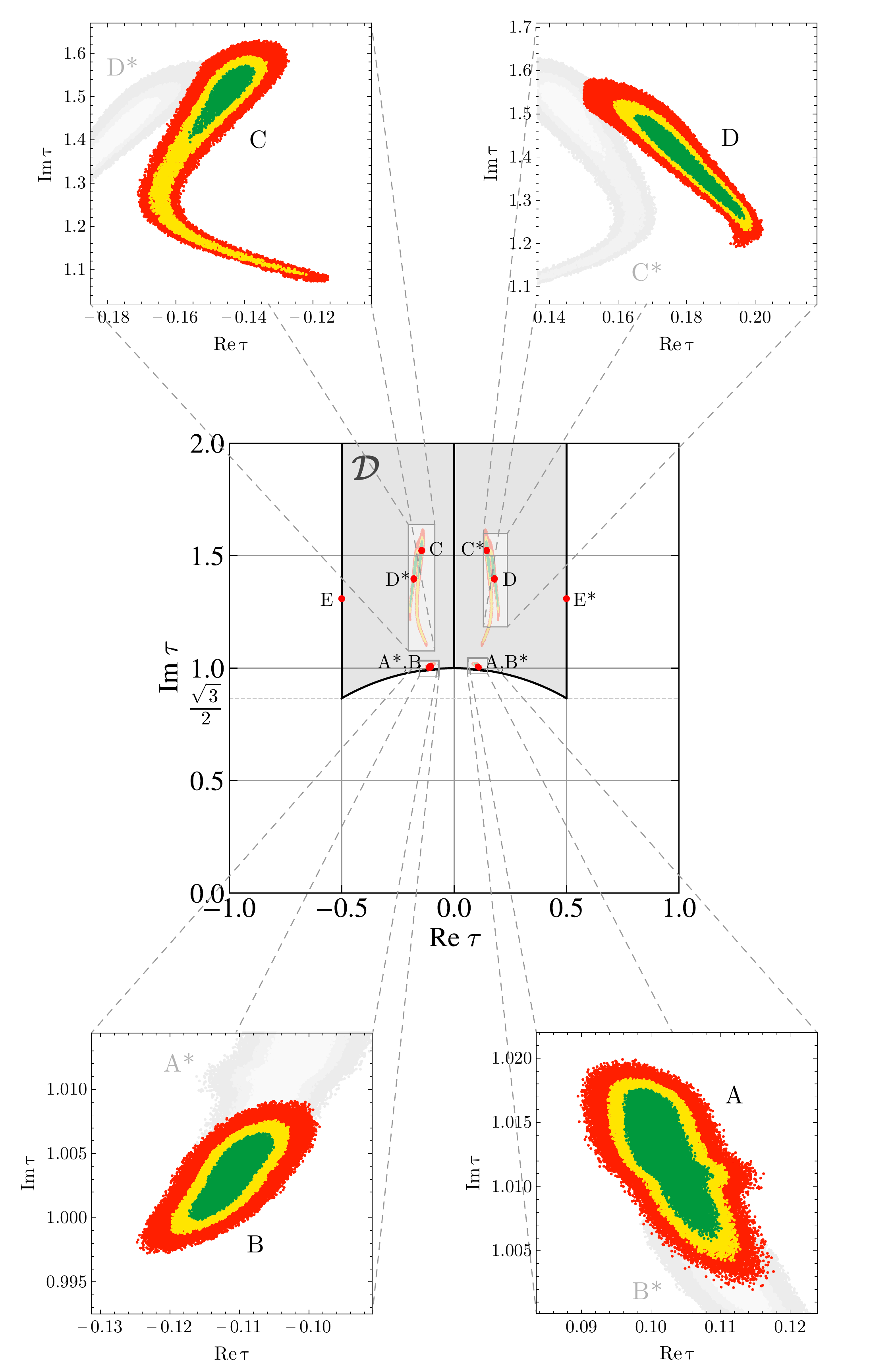}
\caption{
Red dots signal \(\tau\) values inside the fundamental domain
  $\mathcal{D}$ of the modular group corresponding to the
  five pairs of \(\Delta \chi^2\) minima in the case of \((k_{\Lambda}, k_g) = (0, 2)\).
  Here and in the following plots, the green, yellow and red
  regions correspond to 
  $2\s$, $3\s$ and $5\s$ 
  confidence levels, respectively.}
  \label{fig:tau8param}
\end{figure}

 In this case the matrices are given by
eqs.~\eqref{eq:MkL0}, \eqref{eq:ykg2eq} or \eqref{eq:ykg2neq},
and \eqref{eq:lambda1}.
Through numerical search, we find five pairs of distinct local minima of
\(\Delta \chi^2\)
corresponding to five pairs of distinct values of $\tau$. 
The two minima in each pair lead to 
opposite values of the Dirac 
and Majorana phases, but
the same values of all other observables. 
We denote the cases belonging to the first pair as A and A$^*$, 
to the second pair as B and B$^*$, etc. (see Fig.~\ref{fig:tau8param}).
For cases A$^{(*)}$ and B$^{(*)}$ one has $\rho_N \neq \rho_L$, while for the remaining cases $\rho_N = \rho_L$.
Note that starred cases correspond to predictions for $\delta$ not in line with its experimentally allowed $3\sigma$ range.
We present the best fit values along with $2\sigma$ and $3\sigma$
confidence intervals in 
Tables~\ref{tab:case02p8}\,--\,\ref{tab:case02p3}.

 Interestingly, from Fig.~\ref{fig:tau8param} we observe that
6 out of 10 values of \(\tau\) corresponding to local minima lie almost on the boundary of the fundamental domain \(\mathcal{D}\).
The four points which are relatively far from the boundary (C, C$^*$, D, and D$^*$)
correspond to inverted ordering.

The structure of a scalar potential $V$
for the modulus field $\tau$ has been previously 
studied in the context of string compactifications and supergravity 
(see, e.g., \cite{Ferrara:1990ei,Font:1990nt,Cvetic:1991qm}). In Ref.~\cite{Cvetic:1991qm}, considering 
the most general non-perturbative 
effective $\mathcal{N} = 1$ supergravity action in four dimensions,
invariant under modular symmetry, it has been conjectured 
that all extrema of $V$ lie on the boundary
of $\mathcal{D}$ and on the imaginary axis ($\re\t = 0$). 
This conjecture has been checked there in several examples.
If --- as suggested by global analyses --- it turns out that the normal 
ordering of light neutrino masses is realised in Nature, this could be 
considered as an additional indication in favour of the modular 
symmetry approach to flavour.

The models with $(k_\L,k_g) = (0,2)$ analysed by us are 
characterised by six real parameters, $v_d\,\alpha$, $\beta/\alpha$, 
$\gamma/\alpha$, 
$v^2_u\, g^2/\Lambda$, $|g'/g|$, $\im \tau$,
and two phases 
${\rm arg}(g'/g)$ and $\re \tau$.%
\footnote{Notice that the dependence on $\tau$ arises through powers of $\exp(2 \pi i \tau/4)$.}
The three real parameters  $v_d\,\alpha$, $\beta/\alpha$ and 
$\gamma/\alpha$ are fixed by fitting the three values of 
the charged lepton masses.
The remaining three real parameters and two phases 
($v^2_u\, g^2/\Lambda$, $|g^\prime/g|$, $\im \tau$, ${\rm arg}(g^\prime/g)$, 
$\re \tau$)
are used to describe the three neutrino masses,
three neutrino mixing angles 
and the one Dirac and two Majorana
CPV phases present in the PMNS matrix. Obviously, 
the values of some of these altogether nine observables are expected to be correlated.

In the analysis of the five different pairs of models, 
A and A$^*$, B and B$^*,\ldots,$ E and E$^*$, as indicated earlier,
we used as input the $e$, $\mu$ and $\tau$ masses,
the one-dimensional $\chi^2$ projections 
for $\sin^2\theta_{ij}$ from Ref.~\cite{Capozzi:2018ubv} 
and the Gaussian approximation for $\delta m^2$ and $\Delta m^2$.
As a result of the analysis we obtain:
\begin{enumerate}[i)]
\item the best fit values and the $2\sigma$ and $3\sigma$
ranges of ${\rm Re} \tau$,  ${\rm Im} \tau$, 
$\beta/\alpha$, $\gamma/\alpha$, $v_d\,\alpha$,
 $\re (g^\prime/g)$,  $\im (g^\prime/g)$, 
$v^2_u\, g^2/\Lambda$, for which we have 
a sufficiently good quality 
of the fit to the data,
\item the best fit values and the $2\sigma$ and $3\sigma$
allowed ranges of $\sin^2\theta_{ij}$, $\delta m^2$ and $\Delta m^2$, 
to be compared with those found in Ref.~\cite{Capozzi:2018ubv} 
and quoted in Table \ref{tab:globalFit},
\item the {\it predicted} best fit values and 
the $2\sigma$ and $3\sigma$ ranges of the absolute neutrino mass scale ${\rm min}(m_j)$, $j=1,2,3$, 
and of the CPV phases $\delta$, $\alpha_{21}$ and $\alpha_{31}$.
Together with the results on $\delta m^2$, $\Delta m^2$,  
$\sin^2\theta_{12}$ and  $\sin^2\theta_{13}$, this allows us to 
obtain predictions for the sum of neutrino masses 
$\sum_i m_i$ and for the effective Majorana mass in 
neutrinoless double beta decay $|\langle m\rangle|$ (see, e.g., \cite{PDG2018,Bilenky:1987ty}).
\end{enumerate}
These results are reported in Tables~\ref{tab:case02p8}\,--\,\ref{tab:case02p3}. 

\begin{figure}[!ht]
\centering
\includegraphics[width=\textwidth]{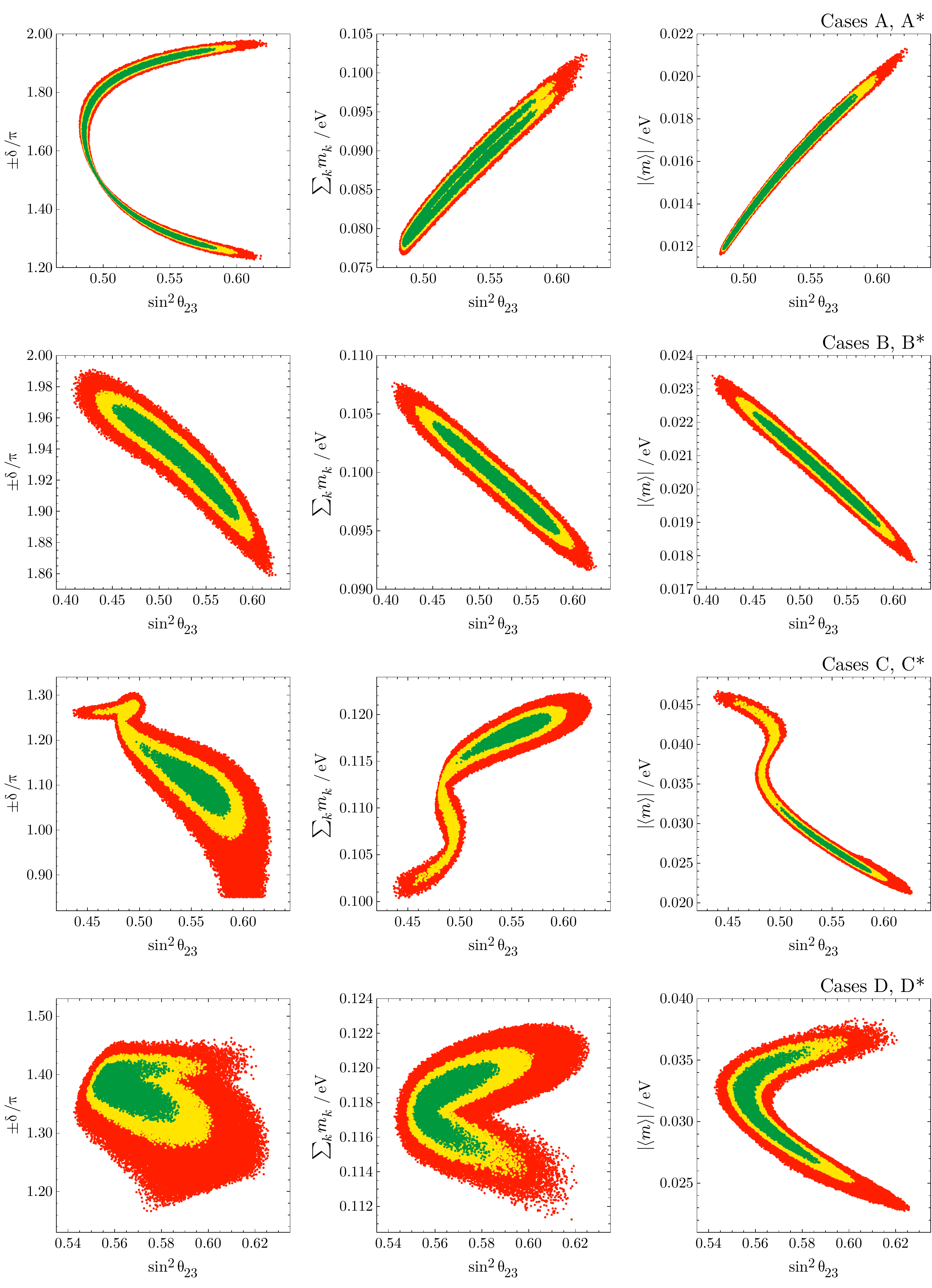}
  \caption{Correlations between $\sin^2\theta_{23}$ and the Dirac CPV phase, the sum of neutrino masses, and the effective Majorana mass in neutrinoless double beta decay, in models with \((k_{\Lambda}, k_g) = (0, 2)\). The plus (minus) sign of $\delta$ refers to the case without (with) an asterisk. 
}
  \label{fig:p8}
\end{figure}

A successful description of the data in the lepton sector, as our analyses show, implies a correlation between the values of $\re\tau$ and $\im\tau$ 
(see Fig.~\ref{fig:tau8param}),
as well as between the values of 
$\im (g^\prime/g)$ and $\im\tau$ 
and of $\re(g^\prime/g)$ and $\im\tau$ 
(see Fig.~\ref{fig:peculiarRegions}). 
In what concerns the neutrino masses and mixing observables,
we find that the value of $\sin^2\theta_{23}$ is correlated 
with the values i) of the Dirac phase $\delta$, 
ii) of $\sum_i m_i$ and iii) of $|\langle m\rangle|$.
These correlations are illustrated in Fig.~\ref{fig:p8}.%
\footnote{In Fig.~\ref{fig:p8} we do not show correlations 
in the cases of the models  E and E$^*$ since 
these models are noticeably less favoured by the data 
than the other four pairs of models 
(see Tables~\ref{tab:case02p8}\,--\,\ref{tab:case02p3}).}
We note that the correlation between the values of 
$\sin^2\theta_{23}$ and   $|\langle m\rangle|$ 
is a consequence, in particular,   
of the correlations between the values of 
$\sin^2\theta_{23}$ and of the Majorana phases 
$\alpha_{21}$ and $\alpha_{31}$.%
\footnote{As a consequence of their 
correlations with $\sin^2\theta_{23}$, the values of 
$\delta$ and of $\alpha_{21}$ and of $\alpha_{31}$ 
are also correlated.}

 Finally, we comment in Appendix~\ref{app:Weinberg}
on the correspondence 
of models with $(k_\L,k_g) = (0,2)$ to the model 
with $k_L = 2$ considered in \cite{Penedo:2018nmg}, where the light neutrino masses are generated via the Weinberg operator.

\subsection{Models with 
\texorpdfstring{$(k_\Lambda,k_g) = (2,2)$}{(kΛ,kg)=(2,2)}}
 In this case the matrices are given by eqs.~\eqref{eq:MkL2}, \eqref{eq:ykg2eq} or \eqref{eq:ykg2neq}, and \eqref{eq:lambda1}.
According to our numerical search, this model cannot accommodate the experimental data.
The best points we have found through numerical search are presented in Table~\ref{tab:case22predictions}.
\begin{table}
\renewcommand{\thetable}{\arabic{table}a}
\centering
\renewcommand{\arraystretch}{1.2}
\begin{tabular}{c|ccc}
  \toprule
  & Best fit value & $2\sigma$ range & $3\sigma$ range \\
  \midrule
  $\re \tau$ & $\pm 0.1045$ & $\pm (0.09597 - 0.1101)$ & $\pm (0.09378 - 0.1128)$ \\
  $\im \tau$ & 1.01 & $1.006 - 1.018$ & $1.004 - 1.018$ \\
  $\beta/\alpha$ & 9.465 & $8.247 - 11.14$ & $7.693 - 12.39$ \\
  $\gamma/\alpha$ & 0.002205 & $0.002032 - 0.002382$ & $0.001941 - 0.002472$ \\
  $\re g'/g$ & 0.233 & $-0.02383 - 0.387$ & $-0.02544 - 0.4417$ \\
  $\im g'/g$ & $\pm 0.4924$ & $\pm (-0.592 - 0.5587)$ & $\pm (-0.6046 - 0.5751)$ \\
  $v_d\,\alpha$ [MeV] & 53.19 \\
  $v_u^2\, g^2 / \Lambda$ [eV] & 0.00933 \\
  \midrule
  $m_e/m_{\mu}$ & 0.004802 & $0.004418 - 0.005178$ & $0.00422 - 0.005383$ \\
  $m_{\mu} / m_{\tau}$ & 0.0565 & $0.048 - 0.06494$ & $0.04317 - 0.06961$ \\
  $r$ & 0.02989 & $0.02836 - 0.03148$ & $0.02759 - 0.03224$ \\
  $\delta m^2$ [$10^{-5} \text{ eV}^2$] & 7.339 & $7.074 - 7.596$ & $6.935 - 7.712$ \\
  $|\Delta m^2|$ [$10^{-3} \text{ eV}^2$] & 2.455 & $2.413 - 2.494$ & $2.392 - 2.513$ \\
  $\sin^2 \theta_{12}$ & 0.305 & $0.2795 - 0.3313$ & $0.2656 - 0.3449$ \\
  $\sin^2 \theta_{13}$ & 0.02125 & $0.01988 - 0.02298$ & $0.01912 - 0.02383$ \\
  $\sin^2 \theta_{23}$ & 0.551 & $0.4846 - 0.5846$ & $0.4838 - 0.5999$ \\
  \midrule
  Ordering & NO \\
  $m_1$ [eV] & 0.01746 & $0.01196 - 0.02045$ & $0.01185 - 0.02143$ \\
  $m_2$ [eV] & 0.01945 & $0.01477 - 0.02216$ & $0.01473 - 0.02307$ \\
  $m_3$ [eV] & 0.05288 & $0.05099 - 0.05405$ & $0.05075 - 0.05452$ \\
  $\textstyle \sum_i m_i$ [eV] & 0.0898 & $0.07774 - 0.09661$ & $0.07735 - 0.09887$ \\
  $|\langle m\rangle|$ [eV] & 0.01699 & $0.01188 - 0.01917$ & $0.01177 - 0.02002$ \\
  $\delta/\pi$ & $\pm 1.314$ & $\pm (1.266 - 1.95)$ & $\pm (1.249 - 1.961)$ \\
  $\alpha_{21}/\pi$ & $\pm 0.302$ & $\pm (0.2821 - 0.3612)$ & $\pm (0.2748 - 0.3708)$ \\
  $\alpha_{31}/\pi$ & $\pm 0.8716$ & $\pm (0.8162 - 1.617)$ & $\pm (0.7973 - 1.635)$ \\
  \midrule
  $N \sigma$ & 0.02005 \\
  \bottomrule
\end{tabular}
\caption{
Best fit values along with $2\sigma$ and $3\sigma$ ranges 
of the parameters and observables 
in cases A and A$^*$, which refer to 
\((k_{\Lambda}, k_g) = (0, 2)\) and 
to a certain region in the $\t$ plane (see Fig.~\ref{fig:tau8param}).}
\label{tab:case02p8}
\end{table}
%
%
%
\begin{table}
\addtocounter{table}{-1}
\renewcommand{\thetable}{\arabic{table}b}
\centering
\renewcommand{\arraystretch}{1.2}
\begin{tabular}{c|ccc}
  \toprule
  & Best fit value & $2\sigma$ range & $3\sigma$ range \\
  \midrule
  $\re \tau$ & $\mp 0.109$ & $\mp (0.1051 - 0.1172)$ & $\mp (0.103 - 0.1197)$ \\
  $\im \tau$ & 1.005 & $0.9998 - 1.007$ & $0.9988 - 1.008$ \\
  $\beta/\alpha$ & 0.03306 & $0.02799 - 0.03811$ & $0.02529 - 0.04074$ \\
  $\gamma/\alpha$ & 0.0001307 & $0.0001091 - 0.0001538$ & $0.0000982 - 0.0001663$ \\
  $\re g'/g$ & 0.4097 & $0.3513 - 0.5714$ & $0.3241 - 0.5989$ \\
  $\im g'/g$ & $\mp 0.5745$ & $\mp (0.5557 - 0.5932)$ & $\mp (0.5436 - 0.5944)$ \\
  $v_d\, \alpha$ [MeV] & 893.2 \\
  $v_u^2\, g^2 / \Lambda$ [eV]& 0.008028 \\
  \midrule
  $m_e/m_{\mu}$ & 0.004802 & $0.004425 - 0.005175$ & $0.004211 - 0.005384$ \\
  $m_{\mu} / m_{\tau}$ & 0.05649 & $0.04785 - 0.06506$ & $0.04318 - 0.06962$ \\
  $r$ & 0.0299 & $0.02838 - 0.03144$ & $0.02757 - 0.03223$ \\
  $\delta m^2$ [$10^{-5} \text{ eV}^2$] & 7.34 & $7.078 - 7.59$ & $6.932 - 7.71$ \\
  $|\Delta m^2|$ [$10^{-3} \text{ eV}^2$] & 2.455 & $2.414 - 2.494$ & $2.393 - 2.514$ \\
  $\sin^2 \theta_{12}$ & 0.305 & $0.2795 - 0.3314$ & $0.2662 - 0.3455$ \\
  $\sin^2 \theta_{13}$ & 0.02125 & $0.0199 - 0.02302$ & $0.01914 - 0.02383$ \\
  $\sin^2 \theta_{23}$ & 0.551 & $0.4503 - 0.5852$ & $0.4322 - 0.601$ \\
  \midrule
  Ordering & NO \\
  $m_1$ [eV] & 0.02074 & $0.01969 - 0.02374$ & $0.01918 - 0.02428$ \\
  $m_2$ [eV] & 0.02244 & $0.02148 - 0.02522$ & $0.02101 - 0.02574$ \\
  $m_3$ [eV] & 0.05406 & $0.05345 - 0.05541$ & $0.05314 - 0.05577$ \\
  $\textstyle \sum_i m_i$ [eV] & 0.09724 & $0.09473 - 0.1043$ & $0.0935 - 0.1056$ \\
  $|\langle m\rangle|$ [eV] & 0.01983 & $0.01889 - 0.02229$ & $0.01847 - 0.02275$ \\
  $\delta/\pi$ & $\pm 1.919$ & $\pm (1.895 - 1.968)$ & $\pm (1.882 - 1.977)$ \\
  $\alpha_{21}/\pi$ & $\pm 1.704$ & $\pm (1.689 - 1.716)$ & $\pm (1.681 - 1.722)$ \\
  $\alpha_{31}/\pi$ & $\pm 1.539$ & $\pm (1.502 - 1.605)$ & $\pm (1.484 - 1.618)$ \\
  \midrule
  $N \sigma$ & 0.02435 \\
  \bottomrule
\end{tabular}
\caption{Best fit values along with $2\sigma$ and $3\sigma$ ranges 
of the parameters and observables 
in cases B and B$^*$, which refer to 
\((k_{\Lambda}, k_g) = (0, 2)\) and 
to a certain region in the $\t$ plane (see Fig.~\ref{fig:tau8param}).}
\label{tab:case02p6}
\end{table}
%
%
%
\begin{table}
\addtocounter{table}{-1}
\renewcommand{\thetable}{\arabic{table}c}
\centering
\renewcommand{\arraystretch}{1.2}
\begin{tabular}{c|ccc}
  \toprule
  & Best fit value & $2\sigma$ range & $3\sigma$ range \\
  \midrule
  $\re \tau$ & $\mp 0.1435$ & $\mp (0.137 - 0.1615)$ & $\mp (0.1222 - 0.168)$ \\
  $\im \tau$ & 1.523 & $1.147 - 1.572$ & $1.088 - 1.594$ \\
  $\beta/\alpha$ & 17.82 & $10.99 - 21.38$ & $9.32 - 23.66$ \\
  $\gamma/\alpha$ & 0.003243 & $0.002518 - 0.003565$ & $0.00227 - 0.003733$ \\
  $\re g'/g$ & $-0.8714$ & $-(0.8209 - 1.132)$ & $-(0.7956 - 1.148)$ \\
  $\im g'/g$ & $\mp 2.094$ & $\mp (1.439 - 2.157)$ & $\mp (1.409 - 2.182)$ \\
  $v_d\, \alpha$ [MeV] & 71.26 \\
  $v_u^2\, g^2 / \Lambda$ [eV]& 0.008173 \\
  \midrule
  $m_e/m_{\mu}$ & 0.004797 & $0.00442 - 0.005183$ & $0.004215 - 0.005378$ \\
  $m_{\mu} / m_{\tau}$ & 0.05655 & $0.04806 - 0.06507$ & $0.04348 - 0.0698$ \\
  $r$ & 0.0301 & $0.02857 - 0.03162$ & $0.0278 - 0.03246$ \\
  $\delta m^2$ [$10^{-5} \text{ eV}^2$] & 7.346 & $7.084 - 7.589$ & $6.946 - 7.717$ \\
  $|\Delta m^2|$ [$10^{-3} \text{ eV}^2$] & 2.44 & $2.4 - 2.479$ & $2.377 - 2.498$ \\
  $\sin^2 \theta_{12}$ & 0.303 & $0.278 - 0.3288$ & $0.2657 - 0.3436$ \\
  $\sin^2 \theta_{13}$ & 0.02175 & $0.02035 - 0.0234$ & $0.01957 - 0.0242$ \\
  $\sin^2 \theta_{23}$ & 0.5571 & $0.4905 - 0.588$ & $0.4551 - 0.6026$ \\
  \midrule
  Ordering & IO \\
  $m_1$ [eV] & 0.0513 & $0.04938 - 0.0518$ & $0.04882 - 0.05207$ \\
  $m_2$ [eV] & 0.05201 & $0.05012 - 0.05248$ & $0.04958 - 0.05274$ \\
  $m_3$ [eV] & 0.01512 & $0.00576 - 0.01594$ & $0.00316 - 0.0163$ \\
  $\textstyle \sum_i m_i$ [eV] & 0.1184 & $0.1053 - 0.1201$ & $0.102 - 0.1208$ \\
  $|\langle m\rangle|$ [eV] & 0.0263 & $0.0239 - 0.04266$ & $0.02288 - 0.04551$ \\
  $\delta/\pi$ & $\pm 1.098$ & $\pm (1.026 - 1.278)$ & $\pm (0.98 - 1.289)$ \\
  $\alpha_{21}/\pi$ & $\pm 1.241$ & $\pm (1.162 - 1.651)$ & $\pm (1.113 - 1.758)$ \\
  $\alpha_{31}/\pi$ & $\pm 0.2487$ & $\pm (0.1474 - 0.3168)$ & $\pm (0.069 - 0.346)$ \\
  \midrule
  $N \sigma$ & 0.0357 \\
  \bottomrule
\end{tabular}
\caption{Best fit values along with $2\sigma$ and $3\sigma$ ranges 
of the parameters and observables 
in cases C and C$^*$, which refer to 
\((k_{\Lambda}, k_g) = (0, 2)\) and 
to a certain region in the $\t$ plane (see Fig.~\ref{fig:tau8param}).}
\label{tab:case02p1b}
\end{table}
%
%
%
\begin{table}
\addtocounter{table}{-1}
\renewcommand{\thetable}{\arabic{table}d}
\centering
\renewcommand{\arraystretch}{1.2}
\begin{tabular}{c|ccc}
  \toprule
  & Best fit value & $2\sigma$ range & $3\sigma$ range \\
  \midrule
  $\re \tau$ & $\pm 0.179$ & $\pm (0.165 - 0.1963)$ & $\pm (0.1589 - 0.199)$ \\
  $\im \tau$ & 1.397 & $1.262 - 1.496$ & $1.236 - 1.529$ \\
  $\beta/\alpha$ & 15.35 & $11.67 - 18.66$ & $10.79 - 21.09$ \\
  $\gamma/\alpha$ & 0.002924 & $0.002582 - 0.003289$ & $0.002443 - 0.003459$ \\
  $\re g'/g$ & $-1.32$ & $-(1.189 - 1.438)$ & $-(1.131 - 1.447)$ \\
  $\im g'/g$ & $\pm 1.733$ & $\pm (1.357 - 1.948)$ & $\pm (1.306 - 2.017)$ \\
  $v_d\,\alpha$ [MeV]& 68.42 \\
  $v_u^2\, g^2 / \Lambda$ [eV] & 0.00893 \\
  \midrule
  $m_e/m_{\mu}$ & 0.004786 & $0.004431 - 0.005186$ & $0.004221 - 0.005386$ \\
  $m_{\mu} / m_{\tau}$ & 0.0554 & $0.0481 - 0.06502$ & $0.04343 - 0.06968$ \\
  $r$ & 0.03023 & $0.02859 - 0.03163$ & $0.02775 - 0.03244$ \\
  $\delta m^2$ [$10^{-5} \text{ eV}^2$] & 7.367 & $7.088 - 7.59$ & $6.937 - 7.713$ \\
  $|\Delta m^2|$ [$10^{-3} \text{ eV}^2$] & 2.437 & $2.4 - 2.479$ & $2.378 - 2.499$ \\
  $\sin^2 \theta_{12}$ & 0.3031 & $0.2791 - 0.3286$ & $0.2657 - 0.3436$ \\
  $\sin^2 \theta_{13}$ & 0.02184 & $0.02038 - 0.02337$ & $0.01954 - 0.0242$ \\
  $\sin^2 \theta_{23}$ & 0.5577 & $0.5509 - 0.5869$ & $0.5482 - 0.6013$ \\
  \midrule
  Ordering & IO \\
  $m_1$ [eV] & 0.05122 & $0.05051 - 0.05185$ & $0.05023 - 0.05212$ \\
  $m_2$ [eV] & 0.05193 & $0.05125 - 0.05253$ & $0.05098 - 0.05279$ \\
  $m_3$ [eV] & 0.01495 & $0.01293 - 0.01613$ & $0.01223 - 0.01649$ \\
  $\textstyle \sum_i m_i$ [eV] & 0.1181 & $0.1149 - 0.1203$ & $0.1139 - 0.1212$ \\
  $|\langle m\rangle|$ [eV] & 0.03104 & $0.02666 - 0.03597$ & $0.02515 - 0.03677$ \\
  $\delta/\pi$ & $\pm 1.384$ & $\pm (1.32 - 1.4245)$ & $\pm (1.271 - 1.437)$ \\
  $\alpha_{21}/\pi$ & $\pm 1.343$ & $\pm (1.227 - 1.457)$ & $\pm (1.171 - 1.479)$ \\
  $\alpha_{31}/\pi$ & $\pm 0.806$ & $\pm (0.561 - 1.092)$ & $\pm (0.448 - 1.149)$ \\
  \midrule
  $N \sigma$ & 0.3811 \\
  \bottomrule
\end{tabular}
\caption{Best fit values along with $2\sigma$ and $3\sigma$ ranges 
of the parameters and observables 
in cases D and D$^*$, which refer to 
\((k_{\Lambda}, k_g) = (0, 2)\) and 
to a certain region in the $\t$ plane (see Fig.~\ref{fig:tau8param}).}
\label{tab:case02p1a}
\end{table}
%
%
%
\begin{table}
\addtocounter{table}{-1}
\renewcommand{\thetable}{\arabic{table}e}
\centering
\renewcommand{\arraystretch}{1.2}
\begin{tabular}{c|cc}
  \toprule
  & Best fit value & $3\sigma$ range \\
  \midrule
  $\re \tau$ & $\mp 0.4996$ & $\mp (0.48 - 0.5084)$ \\
  $\im \tau$ & 1.309 & $1.246 - 1.385$ \\
  $\beta/\alpha$ & 0.000243 & $0.0002004 - 0.0002864$ \\
  $\gamma/\alpha$ & 0.03335 & $0.02799 - 0.03926$ \\
  $\re g'/g$ & $-0.06454$ & $-(0.01697 - 0.1215)$ \\
  $\im g'/g$ & $\mp 0.569$ & $\mp (0.4572 - 0.6564)$ \\
  $v_d\, \alpha$ [MeV] & 1125 \\
  $v_u^2\, g^2 / \Lambda$ [eV] & 0.0174 \\
  \midrule
  $m_e/m_{\mu}$ & 0.004797 & $0.004393 - 0.005197$ \\
  $m_{\mu} / m_{\tau}$ & 0.05626 & $0.04741 - 0.0654$ \\
  $r$ & 0.02985 & $0.02826 - 0.03146$ \\
  $\delta m^2$ [$10^{-5} \text{ eV}^2$] & 7.332 & $7.055 - 7.593$ \\
  $|\Delta m^2|$ [$10^{-3} \text{ eV}^2$] & 2.456 & $2.413 - 2.497$ \\
  $\sin^2 \theta_{12}$ & 0.311 & $0.2895 - 0.3375$ \\
  $\sin^2 \theta_{13}$ & 0.02185 & $0.02041 - 0.02351$ \\
  $\sin^2 \theta_{23}$ & 0.4469 & $0.43 - 0.4614$ \\
  \midrule
  Ordering & NO \\
  $m_1$ [eV] & 0.01774 & $0.01703 - 0.01837$ \\
  $m_2$ [eV] & 0.0197 & $0.01906 - 0.02025$ \\
  $m_3$ [eV] & 0.05299 & $0.05251 - 0.05346$ \\
  $\textstyle \sum_i m_i$ [eV] & 0.09043 & $0.08874 - 0.09195$ \\
  $|\langle m\rangle|$ [eV] & 0.006967 & $0.006482 - 0.007288$ \\
  $\delta/\pi$ & $\pm 1.601$ & $\pm (1.287 - 1.828)$ \\
  $\alpha_{21}/\pi$ & $\pm 1.093$ & $\pm (0.8593 - 1.178)$ \\
  $\alpha_{31}/\pi$ & $\pm 0.7363$ & $\pm (0.3334 - 0.9643)$ \\
  \midrule
  $N \sigma$ & 2.147 \\
  \bottomrule
\end{tabular}
\caption{Best fit values along with $3\sigma$ ranges 
of the parameters and observables 
in cases E and E$^*$, which refer to 
\((k_{\Lambda}, k_g) = (0, 2)\) and 
to a certain region in the $\t$ plane (see Fig.~\ref{fig:tau8param}).}
\label{tab:case02p3}
\end{table}
%
%
%
\clearpage
\thispagestyle{empty}
%

\clearpage
\thispagestyle{empty}
%
%
%
\begin{table}
\centering
\renewcommand{\arraystretch}{1.2}
\begin{tabular}{c|c|c}
  \toprule
  Subcase & $\rho_L = \rho_N$ & $\rho_L \neq \rho_N$ \\
  \midrule
  $\re \tau$ & $\pm 0.1119$ & $\pm 0.2286$ \\
  $\im \tau$ & 1.458 & 0.9736 \\
  $\beta/\alpha$ & 0.0002667 & 0.003258 \\
  $\gamma/\alpha$ & 0.03676 & 8.267 \\
  $\re g'/g$ & 0.9038  & 1.677 \\
  $\im g'/g$ & $\mp 0.3198$ & $\pm 0.004508$ \\
  $v_d\, \alpha$ [MeV]& 1198 & 49.05 \\
  $v_u^2\, g^2 / \Lambda$ [eV]& 0.0352 & 0.002206 \\
  \midrule
  $m_e/m_{\mu}$ & 0.004799 & 0.0048 \\
  $m_{\mu} / m_{\tau}$ & 0.05661 & 0.05657 \\
  $r$ & 0.02999 & 0.03093 \\
  $\delta m^2$ [$10^{-5} \text{ eV}^2$] & 7.355 & 7.509 \\
  $|\Delta m^2|$ [$10^{-3} \text{ eV}^2$] & 2.453 & 2.428 \\
  $\sin^2 \theta_{12}$ & {\color{red}0.4165} & {\color{red}0.3859} \\
  $\sin^2 \theta_{13}$ & 0.02125 & 0.02175 \\
  $\sin^2 \theta_{23}$ & 0.5624 & {\color{red}0.8239} \\
  \midrule
  Ordering & NO & NO \\
  $m_1$ [eV] & 0.01284 & 0.01027 \\
  $m_2$ [eV] & 0.01544 & 0.01343 \\
  $m_3$ [eV] & 0.05152 & 0.0507 \\
  $\textstyle \sum_i m_i$ [eV] & 0.07979 & 0.0744 \\
  $|\langle m\rangle|$ [eV] & $7.381 \cdot 10^{-8}$ & $7.341 \cdot 10^{-6}$ \\
  $\delta/\pi$ & $\pm 1.705$ & $\pm 1.998$ \\
  $\alpha_{21}/\pi$ & $\pm 0.9838$ & $\pm 0.9992$ \\
  $\alpha_{31}/\pi$ & $\pm 0.5056$ & $\pm 0.9989$ \\
  \midrule
  $N \sigma$ & 6.68 & 16.44 \\
  \bottomrule
\end{tabular}
\caption{Best fit values of the parameters and observables 
in the models with \((k_{\Lambda}, k_g) = (2, 2)\).}
\label{tab:case22predictions}
\end{table}
%

\clearpage
\section{Residual Symmetries}
\label{sec:ressymm}
 Residual symmetries arise whenever the VEV of the modulus $\tau$ breaks the modular group $\overline{\Gamma}$ only partially, i.e., the little group (stabiliser) of $\vev{\tau}$ is non-trivial.
There are only 2 inequivalent finite%
\footnote{Note that $\vev{\tau} = i \infty$ breaks $\G_4$ to $\mathbb{Z}_4^T=\{I,T,T^2,T^3\}$.}
points with non-trivial little groups,
namely $\vev{\tau} = -\,1/2 +i\,\sqrt{3}/2  \equiv \tau_L$ (``the left cusp'') with residual symmetry $\mathbb{Z}^{ST}_3 = \{I,ST,(ST)^2\}$, and $\vev{\tau} = i \equiv \tau_C$ with residual symmetry $\mathbb{Z}^{S}_2 = \{I,S\}$ (see, e.g.,~\cite{Serre:1973cs}).
Indeed, the actions of $ST$ on $\tau_L$ and of 
$S$ on $\tau_C$ leave respectively $\tau_L$ and 
$\tau_C$ unchanged.
Any other point with non-trivial little group is related to $\tau_L$ or $\tau_C$ by a modular transformation, and is therefore physically equivalent to it.
For example, $\tau_R = +\,1/2 +i\,\sqrt{3}/2$ (``the right cusp'') has residual symmetry $\mathbb{Z}_3^{TS} = \{I,TS,(TS)^2\}$, and it is related to $\tau_L$ by a $T$ transformation: $\tau_R = T\, \tau_L$.
With one modulus field $\tau$
we can have either the $\mathbb{Z}_3$ or the  $\mathbb{Z}_2$
residual symmetry, and it will be a common symmetry of the charged lepton 
and neutrino sectors of the theory.

In the basis we have employed (see also Appendix~\ref{app:basis}), the triplet irreps of the generators $S$ and $T$ 
have the form:
\be
S =\,\pm\, \frac{1}{3}  
\begin{pmatrix}
-1 & 2\omega^2 & 2\omega \\
2\omega & 2 & -\omega^2 \\
2\omega^2 & -\omega & 2
\end{pmatrix}\,,~~~~
T =\,\pm\,\frac{1}{3}  
\begin{pmatrix}
-1 & 2\omega & 2\omega^2 \\
2\omega & 2\omega^2 & -1 \\
2\omega^2 & -1 & 2\omega
\end{pmatrix}\,,
\label{eq:SandT}
\ee
%
where the plus (minus) sign corresponds to 
the representation $\3$ (representation $\tp$)
of $S_4$. It follows from eq. (\ref{eq:SandT}) that 
in the basis we are using the product of the triplet 
representations of $S$ and $T$ generators is 
a diagonal matrix given by:%
\footnote{The form we get in the triplet representation 
of $ST$ coincides with the form of the triplet representation 
of the $S_4$ generator $T$ in a different presentation for the 
$S_4$ generators (see, e.g., \cite{Petcov:2017ggy} and Appendix~\ref{app:TBM}).}
\be
ST =  
\begin{pmatrix}
1 & 0 & 0 \\
0 & \omega^2 & 0 \\
0 & 0 & \omega
\end{pmatrix}\,.
\label{eq:ST}
\ee
%
In the left cusp point $\vev{\tau} = \tau_L$, 
corresponding to the residual symmetry  $\mathbb{Z}^{ST}_3$,
the five independent modular forms take the following values:
\begin{equation}
\begin{gathered}
 Y_1 = 0\,,~~Y_3 = 0\,,~~Y_5 = 0\,,\\
  Y_2 = i\,2.11219\,,~~Y_4 = -\,i\,2.43895\,,~~
Y_2/Y_4 = -\,\frac{\sqrt{3}}{2}\,.
\label{eq:YjLcusp}
\end{gathered}
\end{equation}
%

 In the point $\vev{\tau} = \tau_C$, invariant under the action of the 
$S$ generator and in which we have the residual symmetry 
$\mathbb{Z}^{S}_2$, the modular forms $Y_2$, $Y_3$, $Y_4$ and $Y_5$
can be expressed in terms of the form $Y_1$:
\begin{equation}
  \begin{gathered}
    Y_2 = -\,\omega^2\,Y_1\,,~~~Y_3 = \frac{2}{3\sqrt{3}}\,\omega\, Y_1\,,\\ 
    Y_4 = \frac{2}{3\sqrt{3}}(1 + \sqrt{6})\,\omega^2\, Y_1\,,~~
    Y_5 = \frac{2}{3\sqrt{3}}(1 - \sqrt{6})\, Y_1\,.
  \end{gathered}
  \label{eq:Yjtc}
\end{equation}
%
At  $\vev{\tau} = i~(=\tau_C)$ we have  
 $Y_1(\tau = i) = 0.7107 + i\,1.231$.

 We could not find models with one modulus field $\tau$ and 
residual symmetry $\mathbb{Z}^{ST}_3$ or $\mathbb{Z}^{S}_2$, 
which are phenomenologically viable. 
Since the residual symmetry is the same for both the charged lepton and neutrino mass matrices,\footnote{Namely, $\rho_L(\gamma)^{\dagger} M_e M_e^{\dagger} \rho_{L}(\gamma) = M_e M_e^{\dagger}$ and $\rho_L(\gamma)^T M_{\nu} \rho_{L}(\gamma) = M_{\nu}$, where $\gamma = ST$ or $S$.} the resulting neutrino mixing matrix always contains zeros, which is ruled out by the data.

 We will consider next the case of having two moduli fields 
in the theory~---~one, $\tau^\ell$, responsible via its VEV 
for the breaking of the modular $S_4$ symmetry in the 
charged lepton sector, and a second one, $\tau^\nu$, 
breaking the modular symmetry in the neutrino sector.
This will be done on purely phenomenological grounds:
we will not attempt to construct a model in which 
the discussed possibility is realised; 
we are not even sure such models exist.

We will assume further that we 
have a residual $\mathbb{Z}^{ST}_3$ symmetry in the charged lepton 
sector and a residual  $\mathbb{Z}^{S}_2$ symmetry 
in the neutrino sector.
Under the indicated conditions, 
one of the charged lepton masses vanishes:
the first column of eq.~\eqref{eq:lambda1} is exactly zero at $\tau_L$, which follows immediately from eq.~\eqref{eq:YjLcusp}. 
However, it is possible to 
render all masses non-vanishing from the outset
if we replace the last Yukawa interaction term in eq.~\eqref{eq:CLform1} with a singlet containing modular forms of weight 6:
\be
\a \left(\Ec_1\, L\, Y_\tp\right)_\1 \Hd + 
\b \left(\Ec_2\, L\, Y^{(4)}_\3\right)_\1 \Hd +
\g \left(\Ec_3\, L\, Y^{(6)}_\3\right)_\1 \Hd\,, 
\label{eq:Weti}
\ee
%
where 
\be
Y^{(6)}_\3 = 
\begin{pmatrix}
Y^2_2 Y_4 - Y^2_1 Y_5 \\
Y^2_2 Y_5 - Y^2_1 Y_3 \\
Y^2_2 Y_3 - Y^2_1 Y_4
\end{pmatrix}
\label
{eq:Y63}
\ee
%
is the only modular form triplet of weight 6 transforming in the $\3$ of $S_4$.
In this case we get diagonal $M_e M_e^\dagger$ at $\tau_L$:
\be
M_e M_e^\dagger = 
v_d^2\,\diag\left(\gamma^2\,
|Y^2_2\,Y_4|^2,\,\beta^2\,|Y_2\,Y_4|^2,\,\alpha^2\,|Y_4|^2
\right).
\label{eq:MeZ3}
\ee
%

\begin{figure}[!ht]
\centering
\includegraphics[width=\textwidth]{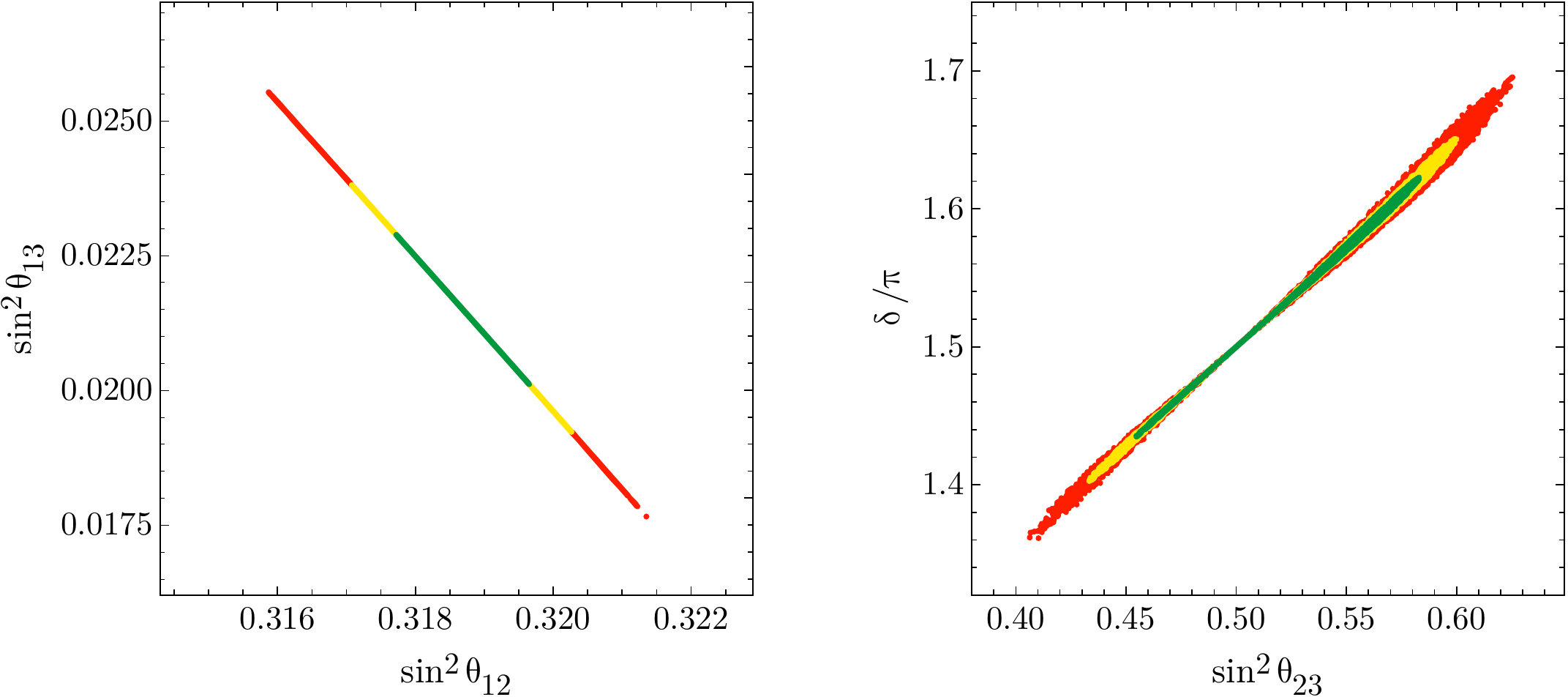}
  \caption{Correlations between $\sin^2\theta_{12}$ and $\sin^2 \theta_{13}$ (left) and between $\sin^2 \theta_{23}$ and $\delta$ (right) in a model with residual symmetry.
}
  \label{fig:symCorr}
\end{figure}

The mixing is therefore determined by the neutrino mass 
matrix having a $\mathbb{Z}_2^S$ symmetry.
It is possible to obtain phenomenologically viable solutions in this scenario. 
For example, in the case $(k_{\Lambda},k_g) = (4,0)$, the neutrino mass matrix 
is given by eqs.~\eqref{eq:MnuSeesaw}, \eqref{eq:MkL4} and \eqref{eq:ykg0}, 
and we find a point 
\begin{equation}
 \tau^{\nu} = i,\,\, \Lambda'/\Lambda = 0.3836 + 1.0894i,\,\, 
\Lambda''/\Lambda = -0.3631 + 0.0039i,
\label{eq:symViable}
\end{equation}
%
consistent with the experimental data at 1$\sigma$ C.L. (for NO spectrum):
\begin{equation}
\begin{gathered}
  r = 0.0299,\quad
  \delta m^2 = 7.34 \cdot 10^{-5} \text{ eV}^2,\quad
  \Delta m^2 = 2.455 \cdot 10^{-3} \text{ eV}^2,\\
  \sin^2 \theta_{12} = 0.3187,\quad
  \sin^2 \theta_{13} = 0.02144,\quad
  \sin^2 \theta_{23} = 0.5512,\\
  m_1 = 0.03437 \text{ eV},\quad
  m_2 = 0.03542 \text{ eV},\quad
  m_3 = 0.0606 \text{ eV},\\
  \textstyle\sum_i m_i = 0.1304 \text{ eV},\quad
  |\langle m\rangle| = 0.0224 \text{ eV},\\
  \delta/\pi = 1.5738,\quad
  \alpha_{21}/\pi = 1.3793,\quad
  \alpha_{31}/\pi = 1.2281.
\end{gathered}
\end{equation}
%
In this case the three masses, three mixing angles and 
three CPV phases in the neutrino sector are described by three 
real parameters, $v^2_u \,g^2/\Lambda$, $|\Lambda'/\Lambda|$ and 
$|\Lambda''/\Lambda|$, and two phases, ${\rm arg}(\Lambda'/\Lambda)$ 
and ${\rm arg}(\Lambda''/\Lambda)$. As a consequence, the values of 
certain neutrino mass and mixing observables should be correlated.
Indeed, through a numerical scan in the vicinity of the point given by eq.~\eqref{eq:symViable} (keeping $\tau^\nu = i$ fixed) we find strong correlations between $\sin^2 \theta_{12}$ and $\sin^2 \theta_{13}$, and between $\sin^2 \theta_{23}$ and $\delta$, as shown in Fig.~\ref{fig:symCorr}.

  We note that instead of considering two different moduli fields, one could also 
realise this scenario with only one modulus $\tau = \tau^{\nu}$ and extra flavon fields.
Below we detail such an alternative model, leading to a diagonal charged lepton mass matrix while preserving at leading order the above results for the neutrino sector.
However, the joint description of  
the lepton and quark flavour, most likely, 
will require the introduction of two different moduli which develop different VEVs.

Suppose that $E^c$ and $L$ are a $\mathbf{3}$ and a $\mathbf{3'}$ triplet, respectively, and the combination $E^c L$ has zero modular weight.
Let us introduce three flavon fields of zero weight, $\phi_{\mathbf{1'}}$, $\phi_{\mathbf{3}}$ and $\phi_{\mathbf{3'}}$, which develop the following VEVs preserving $\mathbb{Z}_3^{ST}$:
\begin{equation}
    \langle \phi_{\mathbf{1'}} \rangle = v_1\,,\quad
    \langle \phi_{\mathbf{3}} \rangle = (v_2, 0, 0)\,,\quad
    \langle \phi_{\mathbf{3'}} \rangle = (v_3, 0, 0)\,.
\label{eq:flavonvevs}
\end{equation}
Let us also assume that the flavon field VEVs are suppressed with respect to
the scale of flavon dynamics $\tilde\Lambda$, $v_i / \tilde\Lambda \ll 1$, so that only the lowest dimension effective operators in the superpotential are relevant.

Since it is impossible to form a trivial singlet from a $\mathbf{3} \otimes \mathbf{3'}$ tensor product, the term $(E^c L)_{\mathbf{1}} H_d$ is not present.
Therefore, the charged lepton mass matrix originates from the linear couplings of $E^c L$ to flavons:
\begin{equation}
    \alpha (E^c L \phi_{\mathbf{1'}})_{\mathbf{1}} H_d +
    \beta (E^c L \phi_{\mathbf{3}})_{\mathbf{1}} H_d +
    \gamma (E^c L \phi_{\mathbf{3'}})_{\mathbf{1}} H_d\,,
\end{equation}
which lead to the following result:
\begin{equation}
    M_e = v_d \,\alpha
    \left[
    \begin{pmatrix}
    1 & 0 & 0 \\
    0 & 0 & 1 \\
    0 & 1 & 0
    \end{pmatrix} +
    \frac{\beta}{\alpha}
    \begin{pmatrix}
    0 & 0 & 0 \\
    0 & 0 & -1 \\
    0 & 1 & 0
    \end{pmatrix} +
    \frac{\gamma}{\alpha}
    \begin{pmatrix}
    2 & 0 & 0 \\
    0 & 0 & -1 \\
    0 & -1 & 0
    \end{pmatrix}
    \right]\,,
\end{equation}
where we have reabsorbed the non-zero VEVs from eq.~\eqref{eq:flavonvevs} into $\alpha$, $\beta$ and $\gamma$.
The relevant product is diagonal:
\begin{equation}
    M_e M_e^\dagger = v_d^2 \,\alpha^2 \diag
    \left(
    \left|1 + 2 \frac{\gamma}{\alpha}\right|^2,\,
    \left|1 - \frac{\beta + \gamma}{\alpha}\right|^2,\,
    \left|1 + \frac{\beta - \gamma}{\alpha}\right|^2
    \right)\,,
\end{equation}
making it possible to fit the charged lepton masses,
by setting $v_d \,\alpha \approx 660$~MeV, $\beta/\alpha \approx 1.34$, and
$1+2\,\gamma/\alpha \approx -7.7 \times 10^{-4}$, with $\alpha \sim \beta \sim \gamma$.
The introduction of the above flavon fields will
imply corrections to the neutrino sector of the theory of the order of $v_i/\tilde\Lambda$.
These can be at the level of a few percent for fairly small $\tan \beta \equiv v_u/ v_d$.

As a final remark, we comment on the famed tri-bimaximal (TBM) mixing~\cite{Harrison:2002er,Harrison:2002kp} 
(see also~\cite{Wolfenstein:1978uw})
in the context of considered residual symmetries in Appendix~\ref{app:TBM}.

\section{Potential Sources of Corrections}
\label{sec:corrections}
One needs to address three potential sources
of corrections, namely, SUSY-breaking effects, the renormalisation group (RG) running,
and corrections to the Kähler potential given in eq.~\eqref{eq:Kahler}. 
The first two effects were analysed in detail in Ref.~\cite{Criado:2018thu}
for closely related modular-invariant models based on the group $A_4$.

As far as SUSY-breaking effects are concerned, the results of Ref.~\cite{Criado:2018thu}
are applicable to the scenario under study. Namely, as demonstrated therein,
corrections to masses and mixing which may not be absorbed in a redefinition of
superpotential parameters can still be made negligible,
provided one realises a sufficient separation between i) the scale $M$ of communication of SUSY-breaking effects to the visible sector and ii) the characteristic scale $m_\mathrm{SUSY} \sim F/M$ of the soft terms,
with $F$ being the spurion VEV assumed to parameterise the breaking of supersymmetry.
Asking for such a gap does not hinder dramatically
the choice of possible values for $m_\mathrm{SUSY}$.

RG effects on neutrino mixing parameters strongly depend on i) $\tan\beta$ and 
ii) the absolute neutrino mass scale $m_\mathrm{min}$. The effects generically become larger when either $\tan\beta$ or $m_\mathrm{min}$ are increased (see, e.g.,~\cite{Antusch:2003kp}). 
Furthermore, for the IO neutrino mass spectrum, 
these effects can be sizeable even for 
$m_\mathrm{min} \rightarrow 0$, 
since in this case the one-loop $\beta$-functions 
for $\theta_{12}$ and $\delta$ are enhanced 
by $\Delta m^2_{23}/\Delta m^2_{21}$ 
independently of $m_\mathrm{min}$ 
(see Table~2 in \cite{Antusch:2003kp}).

It has been found in Ref.~\cite{Criado:2018thu} that for a model 
predicting the normal ordering of neutrinos masses 
with $m_\mathrm{min} \approx 0.01$~eV, 
the RG effects on the predictions 
of the neutrino parameters are negligible 
even for relatively large value of $\tan\beta = 25$. 
For the models considered in our study, 
which lead to the NO spectrum, $m_\mathrm{min} \approx 0.02$~eV 
(0.01~eV) for cases A$^{(*)}$, B$^{(*)}$ and E$^{(*)}$ 
in Tables~\ref{tab:case02p8}, \ref{tab:case02p6} and \ref{tab:case02p3}, respectively 
(for the case characterised by $k_\Lambda = k_g = 2$ in Table~\ref{tab:case22predictions}). 
Thus, we expect the RG corrections to the predictions in Tables~\ref{tab:case02p8}, \ref{tab:case02p6}, \ref{tab:case02p3} and 
\ref{tab:case22predictions} to be negligible. 

 For the second model of Ref.~\cite{Criado:2018thu}, which predicts the IO spectrum, 
it has been shown that for $\tan\beta \lesssim 10$, 
the RG effects are not sizeable.  
It has been also demonstrated that the effects 
depend moderately on the SUSY breaking scale
$m_\mathrm{SUSY}$, with the effects being somewhat less important 
for larger $m_\mathrm{SUSY}$ ($m_\mathrm{SUSY} = 10^4$~GeV and $10^8$~GeV have been compared).
The same conclusions are expected to hold for 
our cases C$^{(*)}$ and D$^{(*)}$ in 
Tables~\ref{tab:case02p1b} and \ref{tab:case02p1a}, respectively, as well as 
for the case characterised by $(k_\Lambda,k_g) = (2,0)$ in Table~\ref{tab:case20predictions}.

 We also would like to note that the case in 
Table~\ref{tab:case20predictions} (without taking the RG effects into account) 
leads to a value of $\sin^2\theta_{12}$ which is
larger than the upper bound of the experimentally allowed 
$3\sigma$ range. If one takes into account the RG evolution 
of the leptonic parameters, 
assuming the predictions in Tables~\ref{tab:case20predictions} to hold 
at the GUT scale, $\Lambda_\mathrm{GUT} \sim 10^{15}-10^{16}$~GeV, 
the situation, in the general case, will worsen. 
The reason for this is the fact that $\sin^2\theta_{12}$ 
increases when running from high to low energies, 
as can be seen in Fig.~2 in \cite{Antusch:2003kp}
(unless the Majorana phase $\alpha_{21} \approx \pi$,
which is not the case in Table~\ref{tab:case20predictions}).

In general, one should also take into account threshold corrections. They depend on the specific SUSY spectrum and, as argued in Ref.~\cite{Criado:2018thu}, 
can be rendered unimportant. This naturally happens if $\tan \beta$ is small.

Finally, modifications to the Kähler potential
can seriously compromise the predictive power of the modular scenario. 
According to Ref.~\cite{Cvetic:1991qm}, there exist
compactifications which do not lead to dangerous instanton contributions to the Kähler potential.
Given the above,
and in consonance with Ref.~\cite{Feruglio:2017spp},
we are taking the simple choice in eq.~\eqref{eq:Kahler}
as a defining pillar of the bottom-up modular scheme.

\section{Summary and Conclusions}
\label{sec:summary}
 In the present article, we have continued 
to develop a new and very interesting 
approach to flavour
proposed in Ref.~\cite{Feruglio:2017spp}.
This approach is based on invariance 
of the physical supersymmetric action 
under the modular group. 
Assuming, in addition, that 
the matter superfields transform 
in irreps of the finite modular group 
$\Gamma_4 \simeq S_4$, we have investigated 
the minimal scenario in which 
the only source of modular symmetry breaking 
is the VEV of the modulus field $\t$ 
and no flavons are introduced. 
Yukawa couplings in such minimal class 
of models are modular forms of level 4, 
transforming in certain irreps of $\G_4$.

Using the basis for 
the lowest non-trivial weight ($k=2$) 
modular forms found 
in Ref.~\cite{Penedo:2018nmg},
we have constructed in a systematic way 
minimal models
in which the light neutrino masses 
are generated via 
the type I seesaw mechanism. 
After stating several simplifying assumptions 
formulated in the beginning of Section~\ref{sec:seesaw}, 
we have classified the minimal models 
according to the weights of the modular forms entering 
i) the Majorana mass-like term
of the gauge singlet neutrinos 
(weight $k_\L$), 
ii) the neutrino Yukawa interaction term 
(weight $k_g$), and 
iii) the charged lepton Yukawa interaction terms 
(weights $k_{\a_i}$, $i=1,2,3$),
 see eq.~\eqref{eq:superpotential}. 
We have shown that the most economic 
(in terms of weights) assignment, 
which yields the correct charged lepton 
mass spectrum, is
$(k_{\a_1},k_{\a_2},k_{\a_3})=(2,4,4)$. 
Adhering to the corresponding matrix 
of charged lepton Yukawa couplings 
given in eq.~\eqref{eq:lambda1}, 
we have demonstrated that in order to have 
a relatively small number of free parameters 
($\leq 8$), both weights $k_\L$ and $k_g$ 
have to be $\leq2$ (Table~\ref{tab:freeparams}).

 Further, we have performed a 
thorough numerical analysis of the models 
with $(k_\L, k_g) = (2,0)$, $(0,2)$ and $(2,2)$.%
\footnote{The weights $(k_\L, k_g) = (0,0)$ 
  lead to a fully degenerate neutrino mass spectrum
  in contradiction with the neutrino oscillation data.} 
We have found that the models characterised by 
$(k_\L, k_g) = (2,0)$ and $(2,2)$ do not 
provide a satisfactory description of 
the neutrino mixing angles 
(Tables~\ref{tab:case20predictions} and~\ref{tab:case22predictions}).
    The models with $(k_\L, k_g) = (0,2)$ instead 
not only successfully accommodate the data on the
charged lepton masses, the neutrino mass-squared differences and the mixing angles, but also 
lead to predictions for the 
absolute neutrino mass scale and 
the Dirac and Majorana CPV phases. 
Our numerical search has revealed 10 
local minima of the $\D\chi^2$ function. 
Each of them is characterised by certain 
values of $\vev{\tau}$
(Fig.~\ref{fig:tau8param})
and other free parameters. 
By investigating regions around these minima 
we have calculated $2\s$ and $3\s$ ranges 
of the observables, which are summarised 
in Tables~\ref{tab:case02p8}\,--\,\ref{tab:case02p3}. Moreover, our numerical procedure 
has shown that the atmospheric mixing parameter 
$\sin^2\th_{23}$ is correlated with 
i) the Dirac CPV phase $\delta$, 
ii) the sum of neutrino masses, and 
iii) the effective Majorana mass in 
neutrinoless double beta decay.
We present these correlations in 
Fig.~\ref{fig:p8}.

 The obtained values of $\vev{\t}$ in the minima 
of the $\D\chi^2$ function lead to a very intriguing observation. 
Namely, 6 of them, which occur  
very close to the boundary of the fundamental 
domain $\mathcal{D}$ of the modular group, 
correspond to the NO neutrino mass spectrum, 
while 4 others, which lie relatively far from the boundary,
correspond to the IO spectrum (Fig.~\ref{fig:tau8param}).
The structure of a scalar potential
for the modulus field $\tau$ has been previously 
studied in the context of string compactifications and supergravity, 
and it has been conjectured in Ref.~\cite{Cvetic:1991qm}
that all extrema of this potential occur on the boundary 
of $\mathcal{D}$ and on the imaginary axis ($\re\t = 0$). 
If --- as suggested by global analyses of 
the neutrino oscillation data --- 
it turns out that the NO spectrum is realised in Nature, this could be considered as an additional indication in favour of the considered modular symmetry approach to flavour.

 Finally, we have performed a residual symmetry 
analysis, based on the fact that the points 
$\vev{\t} = i$, $\vev{\t} = \exp(2\pi i/3)$ and 
$\vev{\t} = \exp(\pi i/3)$ preserve respectively the
$\mathbb{Z}_2^S$, $\mathbb{Z}_3^{ST}$ 
and $\mathbb{Z}_3^{TS}$ subgroups of the 
modular group. While a single preserved 
residual symmetry cannot lead to a viable neutrino mixing matrix, one can assume that 
residual symmetries of the charged lepton 
and neutrino sectors are different. 
In this case, two moduli fields --- one, responsible 
for the breaking of the modular symmetry in the 
charged lepton sector, and a second one 
breaking the modular symmetry in the neutrino sector --- may be needed.
We have considered this scenario on purely phenomenological grounds 
with an assumption of having 
a residual $\mathbb{Z}^{ST}_3$ symmetry in the charged lepton 
sector and a residual $\mathbb{Z}^{S}_2$ symmetry in the neutrino sector.
We have provided a phenomenologically viable example for which 
the charged lepton mass term 
(more specifically, the matrix $M_e\,M_e^\dagger$)
is diagonal, and lepton mixing is fully determined by 
the neutrino mass matrix.%
\footnote{In this example, the weights
 $(k_{\a_1},k_{\a_2},k_{\a_3}) = (2,4,6)$ and $(k_\L,k_g) = (4,0)$. 
 We have pointed out that, alternatively,
 instead of considering two different moduli fields, one could also realise
 this scenario with one modulus $\tau = \tau^{\nu}$ and extra flavon fields. In such a model, the flavons
 develop particularly aligned VEVs which are responsible
 for the diagonal form of the charged lepton mass term.}

 In conclusion, the modular symmetry approach 
to flavour points to a very intriguing 
connection between modular-invariant 
supersymmetric theories 
(possibly originating from string theory) 
and the flavour structures observed at low energies. 
Its predictions will be tested 
with future more precise neutrino oscillation data, with prospective results 
from direct neutrino mass and neutrinoless double beta decay experiments, 
as well as with improved cosmological measurements.

\section*{Acknowledgements}
A.V.T. would like to thank Ferruccio
Feruglio for insightful discussions on 
problems related to this work. 
A.V.T. expresses his gratitude to SISSA and 
the University of Padua, where part of this 
work was carried out, 
for their hospitality and support. 
This project has received funding from the European Union's Horizon 2020 research and innovation programme under the Marie 
Sklodowska-Curie grant agreements No 674896 
(ITN Elusives) and No 690575 (RISE InvisiblesPlus).
This work was supported in part 
by the INFN program on Theore\-tical Astroparticle Physics (P.P.N. and S.T.P.)
and by the  World Premier International Research Center
Initiative (WPI Initiative, MEXT), Japan (S.T.P.).

\appendix
\section{\texorpdfstring{$S_4$ Group Theory}{S4 Group Theory}}

\subsection{Presentation and basis}
\label{app:basis}
 $S_4$ is the symmetric group of permutations of four objects. 
It contains $4! = 24$ elements and admits the five irreps
$\mathbf{1}$, $\mathbf{1'}$, $\mathbf{2}$, $\mathbf{3}$ and $\mathbf{3'}$ 
(see, e.g.,~\cite{Ishimori:2010au}).
While a presentation of $S_4$ in terms of three generators 
(see Appendix~\ref{app:TBM})
is commonly used, 
it proves convenient
to consider in this context a presentation given in terms of two 
generators $S$ and $T$, namely
\begin{align}
S^2 \,=\, (ST)^3 \,=\, T^4 \,=\, I\,.
\label{eq:present42}
\end{align}

Following the identifications described in Ref.~\cite{Penedo:2018nmg},
from the results in
Ref.~\cite{Bazzocchi:2009pv} one 
can find the explicit basis for the $S_4$ generators in different irreps
which we employ in our discussion:
\begin{align}
\mathbf{1}:&\quad \rho(S)= 1 ,\quad \rho(T)= 1\,, \label{eq:irrep1} \\
\mathbf{1'}:&\quad \rho(S)= -1,\quad \rho(T)= -1\,,\\
\mathbf{2}:&\quad \rho(S)= 
\begin{pmatrix}
0 & \omega \\ \omega^2 & 0
\end{pmatrix}
,\quad \rho(T)= 
\begin{pmatrix}
0 & 1 \\ 1 & 0
\end{pmatrix}
\,, \\
\mathbf{3}:&\quad \rho(S)=
\frac{1}{3}
\begin{pmatrix}
-1 & 2 \omega^2 & 2 \omega \\
 2 \omega & 2  & -\omega^2  \\
2 \omega^2 & -\omega & 2  
\end{pmatrix} 
,\quad \rho(T)=
\frac{1}{3}
\begin{pmatrix}
-1 & 2 \omega & 2 \omega^2 \\
 2 \omega & 2 \omega^2 & -1  \\
2 \omega^2 & -1 & 2 \omega  
\end{pmatrix}  \,,\\
\mathbf{3'}:&\quad \rho(S)= 
-\frac{1}{3}
\begin{pmatrix}
-1 & 2 \omega^2 & 2 \omega \\
 2 \omega & 2  & -\omega^2  \\
2 \omega^2 & -\omega & 2  
\end{pmatrix} 
,\quad \rho(T)=
-\frac{1}{3}
\begin{pmatrix}
-1 & 2 \omega & 2 \omega^2 \\
 2 \omega & 2 \omega^2 & -1  \\
2 \omega^2 & -1 & 2 \omega  
\end{pmatrix}\,, \label{eq:irrep3p}
\end{align}
where as usual $\omega = e^{2\pi i/3}$.

\subsection{Clebsch-Gordan coefficients}
\label{app:CGcoefficients} 
For the basis given in the previous subsection,
one can directly make use of the Clebsch-Gordan coefficients listed in Ref.~\cite{Bazzocchi:2009pv},
which we reproduce here for completeness. Entries of each multiplet entering the tensor product are denoted by $\alpha_i$ and $\beta_i$.
\begin{align}
\begin{array}{@{}c@{{}\,\otimes\,{}}c@{{}\,\,=\,\,{}}l@{\quad\sim\quad}l@{}}
\mathbf{1}&\mathbf{r}&\mathbf{r} &\alpha\,\beta_i \\[2mm]
\mathbf{1'}&\mathbf{1'}&\mathbf{1}&\alpha\,\beta \\[2mm]
\mathbf{1'}&\mathbf{2}&\mathbf{2}&
\begin{pmatrix}
\alpha\,\beta_1\\
-\alpha\,\beta_2 
\end{pmatrix}\\[4mm]
\mathbf{1'}&\mathbf{3}&\mathbf{3'}&
\begin{pmatrix}
\alpha\,\beta_1 \\
\alpha\,\beta_2 \\
\alpha\,\beta_3
\end{pmatrix} \\[6mm]
\mathbf{1'}&\mathbf{3'}&\mathbf{3}&
\begin{pmatrix}
\alpha\,\beta_1 \\
\alpha\,\beta_2 \\
\alpha\,\beta_3
\end{pmatrix} 
\end{array}
\end{align}
%
%
%
\begin{align}
\begin{array}{@{}c@{{}\,\otimes\,{}}c@{{}\,\,=\,\,{}}ll@{}}
\mathbf{2}&\mathbf{2}&\mathbf{1}\,\oplus\, \mathbf{1'}\,\oplus\, \mathbf{2} &
\left\{\begin{array}{@{}l@{\quad\sim\quad}l@{}}
\quad \mathbf{1}  & \alpha_1\beta_2+\alpha_2\beta_1\\[2mm]
\quad \mathbf{1'} & \alpha_1\beta_2-\alpha_2\beta_1\\[2mm]
\quad \mathbf{2}  & \begin{pmatrix} \alpha_2\,\beta_2\\
                                    \alpha_1\,\beta_1 
                    \end{pmatrix}
\end{array}\right.
\\[13mm]
\mathbf{2}&\mathbf{3}&\mathbf{3}\,\oplus\,\mathbf{3'}&
\left\{\begin{array}{@{}l@{\quad\sim\quad}l@{}}
\quad \mathbf{3}   & \begin{pmatrix} \alpha_1\,\beta_2 + \alpha_2\,\beta_3 \\
                                     \alpha_1\,\beta_3 + \alpha_2\,\beta_1 \\
                                     \alpha_1\,\beta_1 + \alpha_2\,\beta_2
                     \end{pmatrix}\\[6mm]
\quad \mathbf{3'}  & \begin{pmatrix} \alpha_1\,\beta_2 - \alpha_2\,\beta_3 \\
                                     \alpha_1\,\beta_3 - \alpha_2\,\beta_1 \\
                                     \alpha_1\,\beta_1 - \alpha_2\,\beta_2
                     \end{pmatrix}
\end{array}\right.
\\[16mm]
\mathbf{2}&\mathbf{3'}&\mathbf{3}\,\oplus\,\mathbf{3'}&
\left\{\begin{array}{@{}l@{\quad\sim\quad}l@{}}
\quad \mathbf{3}   & \begin{pmatrix} \alpha_1\,\beta_2 - \alpha_2\,\beta_3 \\
                                     \alpha_1\,\beta_3 - \alpha_2\,\beta_1 \\
                                     \alpha_1\,\beta_1 - \alpha_2\,\beta_2
                     \end{pmatrix}\\[6mm]
\quad \mathbf{3'}  & \begin{pmatrix} \alpha_1\,\beta_2 + \alpha_2\,\beta_3 \\
                                     \alpha_1\,\beta_3 + \alpha_2\,\beta_1 \\
                                     \alpha_1\,\beta_1 + \alpha_2\,\beta_2
                     \end{pmatrix}
\end{array}\right.
\end{array}
\end{align}
%
%
%
\begin{align}
\begin{array}{@{}cll@{}}
\mathbf{3}\,\otimes\,\mathbf{3}\,=\,\mathbf{3'}\,\otimes\,\mathbf{3'}\,=\,\mathbf{1}\,\oplus\, \mathbf{2}\,\oplus\, \mathbf{3}\,\oplus\, \mathbf{3'} &
\left\{\begin{array}{@{}l@{\quad\sim\quad}l@{}}
\quad \mathbf{1}  & \alpha_1\beta_1+\alpha_2\beta_3+\alpha_3\beta_2\\[2mm]
\quad \mathbf{2}  & \begin{pmatrix}
                      \alpha_2\beta_2+\alpha_1\beta_3+\alpha_3\beta_1 \\
                      \alpha_3\beta_3+\alpha_1\beta_2+\alpha_2\beta_1
                    \end{pmatrix}\\[4mm]
\quad \mathbf{3}   & \begin{pmatrix} 
                       2\alpha_1\beta_1-\alpha_2\beta_3-\alpha_3\beta_2 \\
                       2\alpha_3\beta_3-\alpha_1\beta_2-\alpha_2\beta_1 \\
                       2\alpha_2\beta_2-\alpha_1\beta_3-\alpha_3\beta_1
                     \end{pmatrix}\\[6mm]
\quad \mathbf{3'}  & \begin{pmatrix}
                       \alpha_2\beta_3-\alpha_3\beta_2 \\
                       \alpha_1\beta_2-\alpha_2\beta_1 \\
                       \alpha_3\beta_1-\alpha_1\beta_3
                     \end{pmatrix}
\end{array}\right.
\end{array}
\end{align}
%
%
%
\begin{align}
\begin{array}{@{}cll@{}}
\mathbf{3}\,\otimes\,\mathbf{3'}\,=\,\mathbf{1'}\,\oplus\, \mathbf{2}\,\oplus\, \mathbf{3}\,\oplus\, \mathbf{3'} &
\left\{\begin{array}{@{}l@{\quad\sim\quad}l@{}}
\quad \mathbf{1'} & \alpha_1\beta_1+\alpha_2\beta_3+\alpha_3\beta_2\\[2mm]
\quad \mathbf{2}  & \begin{pmatrix}
                      \alpha_2\beta_2+\alpha_1\beta_3+\alpha_3\beta_1 \\
                      -\alpha_3\beta_3-\alpha_1\beta_2-\alpha_2\beta_1
                    \end{pmatrix}\\[4mm]
\quad \mathbf{3}   & \begin{pmatrix} 
                       \alpha_2\beta_3-\alpha_3\beta_2 \\
                       \alpha_1\beta_2-\alpha_2\beta_1 \\
                       \alpha_3\beta_1-\alpha_1\beta_3
                     \end{pmatrix}\\[6mm]
\quad \mathbf{3'}  & \begin{pmatrix}
                       2\alpha_1\beta_1-\alpha_2\beta_3-\alpha_3\beta_2 \\
                       2\alpha_3\beta_3-\alpha_1\beta_2-\alpha_2\beta_1 \\
                       2\alpha_2\beta_2-\alpha_1\beta_3-\alpha_3\beta_1
                     \end{pmatrix}
\end{array}\right.
\end{array}
\end{align}
%

\section{Higher Weight Modular Forms}
\label{app:higherweight}
 In this appendix we present the modular multiplets arising at weights 6, 8 and 10. The linear space of modular forms of weight $k$ (and level $N=4$, corresponding to $\G_4\simeq S_4$) has dimension $2k+1$. At weight $k=6$, one has the irreps: \begin{equation}
\begin{aligned}
Y_\mathbf{1}^{(6)} &= Y_1^3+Y_2^3\,, \quad
Y_{\mathbf{1}'}^{(6)} = Y_1^3-Y_2^3\,,\\
Y_{\mathbf{2}}^{(6)} &= \left(\begin{array}{c}
 Y_1^2 Y_2 \\
 Y_1 Y_2^2
\end{array}\right)\,,\quad
Y_{\mathbf{3}}^{(6)} =
\left(\begin{array}{c}
 Y_2^2 Y_4-Y_1^2 Y_5 \\
 Y_2^2 Y_5-Y_1^2 Y_3 \\
 Y_2^2 Y_3-Y_1^2 Y_4
\end{array}\right)\,,\\
Y_{\mathbf{3}',1}^{(6)} &=
\left(\begin{array}{c}
 Y_1 Y_2 Y_3 \\
 Y_1 Y_2 Y_4 \\
 Y_1 Y_2 Y_5 
\end{array}\right)\,,\quad
Y_{\mathbf{3}',2}^{(6)} =
\left(\begin{array}{c}
 Y_5 Y_1^2+Y_2^2 Y_4 \\
 Y_3 Y_1^2+Y_2^2 Y_5 \\
 Y_4 Y_1^2+Y_2^2 Y_3
\end{array}\right)\,,
\end{aligned}
\end{equation}
corresponding to a total dimension of 13. 
At weight $k=8$ one has
\begin{equation}
\begin{aligned}
Y_\mathbf{1}^{(8)} &= Y_1^2 Y_2^2\,, \quad
Y_{\mathbf{2},1}^{(8)} = \left(\begin{array}{c}
 Y_1 Y_2^3 \\
 Y_1^3 Y_2
\end{array}\right)\,, \quad
Y_{\mathbf{2},2}^{(8)} =
\left(Y_1^3-Y_2^3\right)\left(\begin{array}{c}
 Y_1  \\
 -Y_2
\end{array}\right)\,,\\
Y_{\mathbf{3},1}^{(8)} &=
\left(Y_1^3-Y_2^3\right) \left(\begin{array}{c}
 Y_3 \\
 Y_4 \\
 Y_5
\end{array}\right)\,,\quad
Y_{\mathbf{3},2}^{(8)} =
\left(\begin{array}{c}
 Y_1^2 Y_2 Y_4-Y_1 Y_2^2 Y_5 \\
 Y_1^2 Y_2 Y_5-Y_1 Y_2^2 Y_3 \\
 Y_1^2 Y_2 Y_3-Y_1 Y_2^2 Y_4
\end{array}\right)\,,\\
Y_{\mathbf{3}',1}^{(8)} &=
\left(Y_1^3+Y_2^3\right)\left(\begin{array}{c}
  Y_3 \\
  Y_4 \\
  Y_5
\end{array}\right)\,,\quad
Y_{\mathbf{3}',2}^{(8)} =
\left(\begin{array}{c}
 Y_1^2 Y_2 Y_4+Y_1 Y_2^2 Y_5 \\
 Y_1 Y_2^2 Y_3+Y_1^2 Y_2 Y_5 \\
 Y_1^2 Y_2 Y_3+Y_1 Y_2^2 Y_4
\end{array}\right)\,,
\end{aligned}
\end{equation}
corresponding to a total dimension of 17. 
Finally, at weight $k=10$ one has
\begin{equation}
\begin{aligned}
Y_\mathbf{1}^{(10)} &= Y_2 Y_1^4+Y_2^4 Y_1\,, \quad
Y_{\mathbf{1}'}^{(10)} = Y_1 Y_2 \left(Y_1^3-Y_2^3\right) \,, \\
Y_{\mathbf{2},1}^{(10)} &= \left(\begin{array}{c}
 Y_1^3 Y_2^2 \\
 Y_1^2 Y_2^3
\end{array}\right)\,, \quad
Y_{\mathbf{2},2}^{(10)} =
\left(Y_1^3-Y_2^3\right)\left(\begin{array}{c}
 -Y_2^2  \\
 Y_1^2
\end{array}\right)\,,\\
Y_{\mathbf{3},1}^{(10)} &=
\left(\begin{array}{c}
 Y_1 Y_2^3 Y_4-Y_1^3 Y_2 Y_5 \\
 Y_1 Y_2^3 Y_5-Y_1^3 Y_2 Y_3 \\
 Y_1 Y_2^3 Y_3-Y_1^3 Y_2 Y_4
\end{array}\right)\,,\quad
Y_{\mathbf{3},2}^{(10)} =
\left(Y_1^3-Y_2^3\right)\left(\begin{array}{c}
 Y_1  Y_4+Y_2  Y_5 \\
 Y_2  Y_3+Y_1  Y_5 \\
 Y_1  Y_3+Y_2  Y_4 
\end{array}\right)\,,\\
Y_{\mathbf{3}',1}^{(10)} &=
\left(\begin{array}{c}
 Y_1^2 Y_2^2 Y_3 \\
 Y_1^2 Y_2^2 Y_4 \\
 Y_1^2 Y_2^2 Y_5
\end{array}\right)\,,\quad
Y_{\mathbf{3}',2}^{(10)} =
\left(\begin{array}{c}
 Y_1 Y_2^3 Y_4+Y_1^3 Y_2 Y_5 \\
 Y_1^3 Y_2 Y_3+Y_1 Y_2^3 Y_5 \\
 Y_1 Y_2^3 Y_3+Y_1^3 Y_2 Y_4 
\end{array}\right)\,,\\
Y_{\mathbf{3}',3}^{(10)} &=
\left(Y_1^3-Y_2^3\right)\left(\begin{array}{c}
 Y_1  Y_4-Y_2  Y_5 \\
 Y_1  Y_5-Y_2  Y_3 \\
 Y_1  Y_3-Y_2  Y_4 
\end{array}\right)\,,
\end{aligned}
\end{equation}
corresponding to a total dimension of 21. 
The correct dimensionality of each linear space is guaranteed via an appropriate number of constraints relating products of modular forms.

\section{Numerical Procedure}
\label{app:numerical}
 Our goal is to explore phenomenologically viable regions in the parameter space, i.e.,
\begin{equation}
  \left\{ p_i:~l(p_i) \leq l_{\text{max}}\right\},
  \label{eq:viableRegion}
\end{equation}
where \(l(p_i)\) is the ``loss'' objective function, which we define as \(l(p_i) \equiv N\sigma(p_i) \equiv \sqrt{\Delta \chi^2 (p_i)}\), and \(l_{\text{max}}\) is the threshold, which we set to 3, so that it corresponds to compatibility with the observed data at 3\(\sigma\) confidence level.

We decompose this problem into two parts: first, we find local minima \(p_i^{(1)},\, p_i^{(2)},\, \ldots\) of \(l(p_i)\), and then we explore connected regions around the minima \(p_i^{(n)}\) that satisfy the constraint \(l \left( p_i \right) \leq l_{\text{max}}\).

To find local minima of \(l(p_i)\), we use the following algorithm:
\begin{enumerate}
\item Pick parameters \(p_i\) at random until we find a ``good enough'' point such that \(l(p_i) < l_{0.01}\).
The threshold \(l_{0.01}\) is a 0.01 quantile of the \(l(p_i)\) distribution, i.e., it is chosen in such a way that we accept roughly 1\% points.
We use this preliminary step to filter out unpromising points which are very far from the regions of interest.
Note that typically \(l_{0.01} > l_{\text{max}}\), i.e. the regions of interest cover only a tiny fraction of the parameter space, so this step is needed to speed up the computation.
\item Run a conventional gradient-based local minimisation algorithm for the objective function \(l(p_i)\) starting from this point.
If the resulting local minimum satisfies the constraint \(l \leq l_{max}\), then add it to a set of viable minima.
\item Repeat steps 1 and 2 until we stop finding any new viable minima.
\end{enumerate}

\begin{figure}[t]
  \centering
  \includegraphics[width=\textwidth]{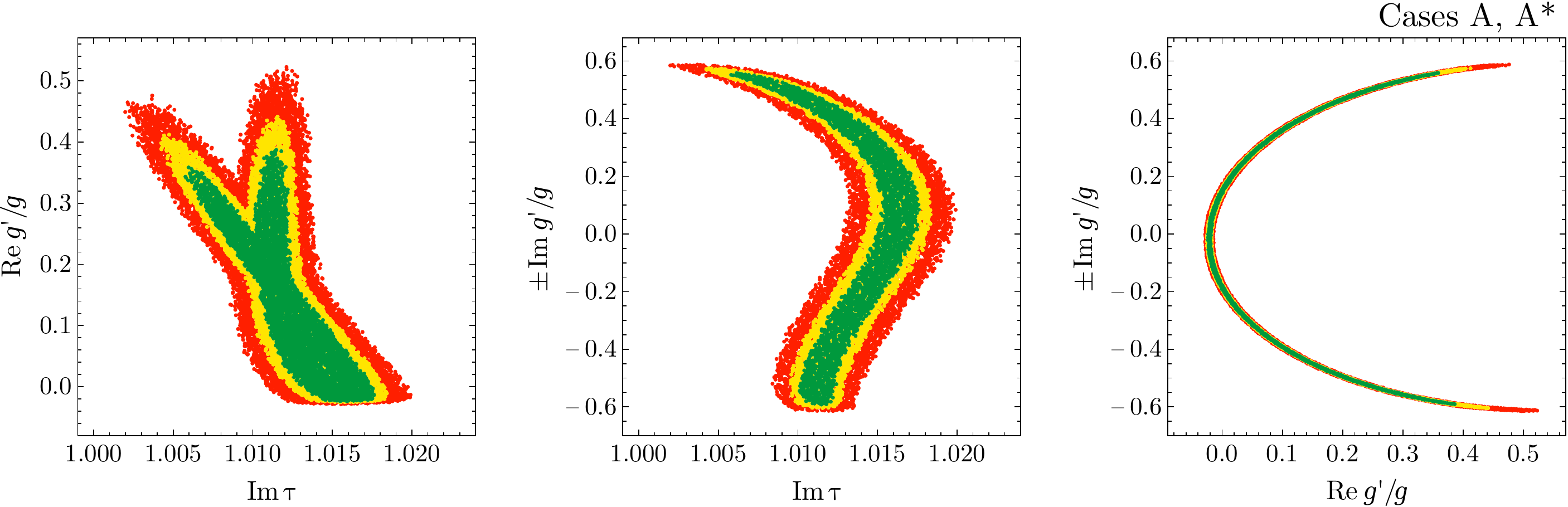}
  \caption{Correlations between the parameters $\im \tau$, $\re g'/g$ and $\im g'/g$
  in cases A and A$^*$, which refer to 
\((k_{\Lambda}, k_g) = (0, 2)\) and 
to a certain region in the $\t$ plane (see Fig.~\ref{fig:tau8param}).
 The plus (minus) signs refer to the case without (with) an asterisk.}
  \label{fig:peculiarRegions}
\end{figure}

At this point, we have a set of distinct viable minima, so for each of them we have to explore the viable region around them.
A simple approach to the problem is to vary parameters \(p_i\) individually until the objective function \(l(p_i)\) increases to \(l_{\text{max}}\).
It corresponds to approximation of the viable region with a parallelepiped.
A more sophisticated approach is to approximate the viable region with an ellipsoid by expanding \(l(p_i)\) around the minimum up to the second order.
However, neither of these approaches work well in our setting due to peculiar shapes of viable regions, see, e.g., Fig.~\ref{fig:peculiarRegions}.
%
Typically, only a small part of a viable region can be approximated with a parallelepiped or an ellipsoid, therefore such approximations lead to a significant underestimation of the full viable parameter space.

Instead, we explore a viable region with a random walk process known as the Metropolis algorithm.
The algorithm mimics the Brownian motion of a probe particle in a potential.
The procedure is as follows:
\begin{enumerate}
\item 
  Define a ``potential'' 
\be
V(p_i) = \begin{cases} l(p_i)\,, \text{ if } l(p_i) \leq l_{\text{max}}'\,, \\ +\infty\,, \text{ otherwise.} \end{cases}
\ee
  We set \(l_{\text{max}}' = 5 > l_{\text{max}}\) in order to make the boundary \(l(p_i) = l_{\text{max}}\) clearly visible in the plots.
\item
  Start a sequence with any point \(p_i^{(0)}\) from the viable region, e.g., the local minimum found previously.
\item
  At iteration $t$, generate a candidate point \(p_i'\) according to a Gaussian distribution centred at \(p_i^{(t)}\) with covariance $\Sigma = \diag(\sigma_1^2, \ldots, \sigma_6^2)$, where $\sigma_i$ are ``step sizes'' along different axes, which have to be tuned.
\item
  Accept the candidate point with probability $\alpha = \min \left[ 1, \exp \left( ({V(p_i^{(t)}) - V(p_i')})/{T} \right) \right]$, where $T$ is the ``temperature'' to be tuned.
\item
  Repeat steps 3 and 4 until the region is fully explored.
\end{enumerate}

One can show that the resulting sequence is distributed according to the Boltzmann (Gibbs) distribution $P(p_i) \propto \exp \left( -V(p_i)/T \right)$, which explains our choice of the potential \(V(p_i)\).

\section{Complex Conjugation of Modular Forms}
\label{app:conjModularForms}
 Suppose that $f_i(\tau)$ is a modular multiplet of weight~$k$ and level~$N$, i.e. $f_i(\tau)$ are holomorphic functions transforming under the modular group~$\overline{\Gamma}$ as given by eq.~\eqref{eq:vvmodforms}.
Let us define $\tilde{f}_i(\tau) \equiv f_i^{\star}(-\tau^{\star})$.
$\tilde{f}_i(\tau)$ are holomorphic and well-defined in the upper half-plane.
Under the modular group they transform as
\begin{equation}
\begin{aligned}
    \tilde{f}_i \left(\frac{a\tau + b}{c\tau + d} \right) &=
    f_i^{\star} \left(-\frac{a\tau^{\star} + b}{c\tau^{\star} + d} \right) =
    f_i^{\star} \left( \frac{a(-\tau^{\star}) + (-b)}{(-c)(-\tau^{\star}) + d} \right) \\
    &= \left[
      \left( (-c)(-\tau^{\star}) + d \right)^{k} \rho_{ij} \left( \begin{pmatrix} a & -b \\ -c & d \end{pmatrix} \right) f_j(-\tau^{\star})
    \right]^{\star} \\
    &= (c\tau + d)^{k} \rho_{ij}^{\star} \left( \begin{pmatrix} a & -b \\ -c & d \end{pmatrix} \right) \tilde{f}_j(\tau)\,.
\end{aligned}  
  \label{eq:conjMultiplet}
\end{equation}
%
Note that $\rho_{ij}^{\star} \left( \left( \begin{smallmatrix} a & -b \\ -c & d \end{smallmatrix} \right) \right)$ is a well-defined unitary representation of $\G_N$, since
\begin{equation}
  \begin{pmatrix} a & b \\ c & d \end{pmatrix} \mapsto \begin{pmatrix} a & -b \\ -c & d \end{pmatrix}
\end{equation}
%
is an automorphism of $\Gamma$ which preserves $\Gamma(N)$.

From eq.~\eqref{eq:conjMultiplet} it follows that $\tilde{f}_i(\tau)$ is a modular multiplet of the same weight, level and dimension as $f_i(\tau)$.
In the case of level 4 and weight 2, there is only one modular multiplet of dimension 2, which is $Y_{\2}(\tau)$, and one modular multiplet of dimension 3, which is $Y_{\tp}(\tau)$, so the conjugated multiplets $Y_{\2}^{\star}(-\tau^{\star})$ and $Y_{\tp}^{\star}(-\tau^{\star})$ should be related by linear transformations to $Y_{\2}(\tau)$ and $Y_{\tp}(\tau)$, respectively.
From the $q$-expansions one can find that these transformations coincide up to a sign with the 
inverse of the modular transformation $T$
in the corresponding representation:
\begin{equation}
  \label{eq:conjLevel4}
  \begin{aligned}
    Y_{\2}^{\star}(-\tau^{\star}) &= -\rho_{\2}\left(T^{-1}\right) Y_{\2}(\tau), \\
    Y_{\tp}^{\star}(-\tau^{\star}) &= -\rho_{\tp}\left(T^{-1}\right) Y_{\tp}(\tau).
  \end{aligned}
\end{equation}
%

\section{Correspondence to the Weinberg Operator Models}
\label{app:Weinberg}
 Reference~\cite{Penedo:2018nmg} studies modular $S_4$ models in which the neutrino masses are generated via the supersymmetric Weinberg operator.
One might expect that seesaw type I modular $S_4$ models at low energies should correspond to a subclass of the Weinberg operator modular $S_4$ models.
However, this is not the case for most choices of modular weights of $N_i^c$.

Namely, in the case $k_N \neq 0 \Leftrightarrow k_{\Lambda} \neq 0$ the mass matrix $M$ of the heavy neutrinos $N^c$ depends explicitly on $\tau$.
The effective light neutrino mass matrix $M_\nu$ obtained by integrating out $N^c$ includes the inverse of $M$ (see eq.~\eqref{eq:MnuSeesaw}), so its entries are polynomials of the modular forms divided by $\det M$.
It is straightforward to check that $\det M$ is a modular singlet of weight $6k_N$ which depends non-trivially on $\tau$.
Moreover, such a modular singlet vanishes for certain values of $\tau$ (see, e.g.,~\cite{Serre:1973cs}).%
\footnote{In this case, it is actually impossible to integrate out all $N_i^c$ for such values of $\tau$, because $\det M = 0$ implies that at least one of the $N_i^c$ fields is massless.}
For example, in the case $k_N = 1 \Leftrightarrow k_\Lambda = 2$ the matrix $M$ is given by eq.~\eqref{eq:MkL2}, and $\det M \propto Y_\1^{(6)} = Y_1^3 + Y_2^3$ vanishes at $\tau = \tau_C = i$ as follows from eq.~\eqref{eq:Yjtc}.
Therefore, the resulting light neutrino mass matrix $M_\nu$ is non-analytic in $\tau$.
On the contrary, the neutrino mass matrix originating from the Weinberg operator without the seesaw mechanism is analytic in $\tau$ by construction.

The only choice of $k_N$ which saves analyticity is $k_N = 0 \Leftrightarrow k_\Lambda = 0$.
Indeed, the mass matrix for the heavy neutrinos is independent of $\tau$ in this case, as can be seen from eq.~\eqref{eq:MkL0}.
Therefore, the heavy neutrinos can be safely integrated out without loss of analyticity, leading to a Weinberg operator modular $S_4$ model.

For the phenomenologically viable case $(k_\L,k_g)=(0,2)$ considered in this article 
and corresponding to $k_L = 2$ (see eq.~\eqref{eq:weights}), 
one can show by direct calculation that the light neutrino mass matrix $M_{\nu}$ coincides with the light neutrino mass matrix from Ref.~\cite{Penedo:2018nmg} in the case $k_L = 2$ (model II therein) with a specific choice of parameters:
\begin{equation}
  \begin{aligned}
    & \begin{cases}
    \left( g/g' \right)_{\text{W}} = 2\, \cfrac{1 - \frac{4}{9} \left( g'/g \right)^2}{1 + \frac{4}{9} \left( g'/g \right)^2}\,,\\[4mm]
    \left( g''/g' \right)_{\text{W}} = 2\, \cfrac{\left( g'/g \right) \left( 1 + \frac{2}{3 \sqrt{3}} \left( g'/g \right) \right)}{1 + \frac{4}{9} \left( g'/g \right)^2}\,,
  \end{cases}
  & \quad \text{for} \quad \rho_N = \rho_L\,,\\[4mm]
  & \begin{cases}
    \left( g/g' \right)_{\text{W}} = -2\,,\\[4mm]
    \left( g''/g' \right)_{\text{W}} = 2\, \cfrac{\left( g'/g \right)}{1 - \frac{2}{\sqrt{3}} \left( g'/g \right)}\,,
  \end{cases}
  & \quad \text{for} \quad \rho_N \neq \rho_L\,.
  \end{aligned}
  \label{eq:seesaw2Weinberg}
\end{equation}
%
Here $g'/g$ is the light neutrino mass matrix parameter as defined in this article in eqs.~\eqref{eq:ykg2eq} and \eqref{eq:ykg2neq}, and $\left( g/g' \right)_{\text{W}}$, $\left( g''/g' \right)_{\text{W}}$ are the corresponding parameters of model II as defined in Ref.~\cite{Penedo:2018nmg}.
Note that the different subcases $\rho_N = \rho_L$ and $\rho_N \neq \rho_L$ translate into two different subspaces of codimension 2 of the parameter space of model II.
Moreover, since the charged lepton mass matrices are the same in these models, they lead to the same observables.
One can check, for instance, that the best fit values presented in Tables~\ref{tab:case02p8}\,--\,\ref{tab:case02p3} can be realised in the Weinberg operator model II by a transformation of parameters according to eq.~\eqref{eq:seesaw2Weinberg}.

\section{Residual Symmetries and Tri-Bimaximal Mixing}
\label{app:TBM}
 It is interesting to check whether 
TBM mixing \cite{Harrison:2002er,Harrison:2002kp} 
(see also \cite{Wolfenstein:1978uw}) can be realised 
with the residual symmetries corresponding to the self-dual points. 
The presentation of $S_4$ 
which naturally leads to the TBM mixing matrix
\be
U_\mathrm{TBM} = 
\begin{pmatrix}
\sqrt{\frac{2}{3}} & \sqrt{\frac{1}{3}} & 0 \\[0.2cm]
-\sqrt{\frac{1}{6}} & \sqrt{\frac{1}{3}} & -\sqrt{\frac{1}{2}} \\[0.2cm]
-\sqrt{\frac{1}{6}} & \sqrt{\frac{1}{3}} & \sqrt{\frac{1}{2}}
\end{pmatrix}
\label{UTBM}
\ee
%
involves three generators $\tilde S$, $\tilde T$ and $\tilde U$ \cite{King:2009mk}, 
satisfying \cite{Hagedorn:2010th}
\be
\tilde S^2 = \tilde T^3 = \tilde U^2 = (\tilde S \tilde T)^3 = (\tilde S \tilde U)^2 = (\tilde T \tilde U)^2 = (\tilde S \tilde T \tilde U)^4 = I\,.
\label{eq:S4presentation}
\ee
%
In the basis for $S_4$ from \cite{King:2009mk} 
the matrices for these three generators
in the irrep $\3$ read
\be
\rho(\tilde S) = \frac{1}{3}
\begin{pmatrix}
-1 & 2 & 2 \\
2 & -1 & 2 \\
2 & 2 & -1
\end{pmatrix}\,,
\quad
\rho(\tilde T) = 
\begin{pmatrix}
1 & 0 & 0 \\
0 & \o^2 & 0 \\
0 & 0 & \o
\end{pmatrix}
\quad
\mathrm{and}
\quad
\rho(\tilde U) = -
\begin{pmatrix}
1 & 0 & 0 \\
0 & 0 & 1 \\
0 & 1 & 0
\end{pmatrix}\,,
\label{eq:S4generators}
\ee
where $\omega = e^{2\pi i/3}$. 
It is worth noting that this presentation of $S_4$ has been worked out in \cite{King:2009mk}
in order to connect $S_4$ downwards to $A_4$ generated by $\tilde S$ and $\tilde T$. 
Indeed, $\rho(\tilde S)$ and $\rho(\tilde T)$ in eq.~\eqref{eq:S4generators} represent 
the widely known Altarelli-Feruglio basis for $A_4$ \cite{Altarelli:2005yx}.
The $\tilde S$ and $\tilde U$ elements generate the $\mathbb{Z}_2^{\tilde S} \times \mathbb{Z}_2^{\tilde U}$ 
subgroup of $S_4$. When preserved in the neutrino sector, 
this subgroup leads to TBM mixing, since $\rho(\tilde S)$ and $\rho(\tilde U)$ 
are simultaneously diagonalised by $U_\mathrm{TBM}$.

 Considering the presentation of $S_4$ in terms of two generators $S$ and $T$ 
given in eq.~\eqref{eq:present42}, which we repeat here for convenience,
\be
S^2 \,=\, (ST)^3 \,=\, T^4 \,=\, I\,,
\label{eq:present42again}
\ee
%
one can show that%
\footnote{We have checked this correspondence for all five irreps of $S_4$, 
using eqs.~\eqref{eq:irrep1}\,--\,\eqref{eq:irrep3p} and 
the corresponding expressions for
$\rho(\tilde S)$, $\rho(\tilde T)$ and $\rho(\tilde U)$ 
from Appendix~A of \cite{Hagedorn:2010th}. 
Note that due to the choice of $\tr(\tilde U)$ made in Ref.~\cite{Hagedorn:2010th}, 
our irrep $\3$ ($\tp$) corresponds to their $\tp$ ($\3$).}
\be
\begin{cases}
S = \tilde S \tilde T \tilde S \tilde U\,, \\[2mm]
T = \tilde T^2 \tilde S \tilde T \tilde U\,,
\end{cases}
\quad \text{or vice versa} \qquad
\begin{cases}
\tilde S = T^2\,, \\[2mm]
\tilde T = S T\,, \\[2mm]
\tilde U = S T^2 S T^3\,.
\end{cases}
\ee
%
Thus, the preserved $S$ generator corresponds to $\tilde S \tilde T \tilde S \tilde U$, 
and a 3-dimensional $\rho(\tilde S \tilde T \tilde S \tilde U)$ is not diagonalised by $U_\mathrm{TBM}$.
In order to preserve $\tilde S$ one would need to preserve $T^2$. 
The value of $\vev{\t} = i\infty$ has residual symmetry
$\mathbb{Z}_4^T = \{I, T, T^2, T^3\}$, which contains $T^2$, 
but this does not work because $\rho(T)$ itself is not diagonalised by $U_\mathrm{TBM}$. 
The question is whether there exists
a value of $\vev{\t}$ which preserves the $\mathbb{Z}_2^{T^2} = \{I, T^2\}$ subgroup of $S_4$. 
There are only two inequivalent finite points 
in the $\vev{\t}$ plane with non-trivial little groups ($\t_L$ and $\t_C$), 
both of which we have considered in Section~\ref{sec:ressymm}, 
and they do not preserve $\mathbb{Z}_2^{T^2}$. 
Thus, it seems rather difficult to realise TBM mixing in the considered set-up.

\bibliography{ModularS4Models}

\end{document}